\documentclass[10pt]{amsart}

\usepackage{graphicx}
\usepackage{dcolumn}
\usepackage{bm}

\usepackage[utf8]{inputenc}
\usepackage[T1]{fontenc}
\usepackage{mathptmx}

\usepackage{amsthm}
\usepackage{amssymb}
\usepackage{eurosym}
\usepackage[colorlinks,linkcolor=blue,citecolor=blue,urlcolor=blue]{hyperref}
\usepackage{caption} 
\newcommand{\beq}{\begin{equation}}
\newcommand{\eeq}{\end{equation}}
\newcommand{\bea}{\begin{eqnarray}}
\newcommand{\eea}{\end{eqnarray}}

\theoremstyle{definition}

\begin{document}


\author[Juan C. King]{Juan C. King}
\address{Centro de Investigaci\'{o}n Operativa, Universidad Miguel Hern\'{a}%
ndez, 03202 Elche, Spain}
\email{juan.king@goumh.umh.es}
\author[Roberto Dale]{Roberto Dale}
\address{Centro de Investigaci\'{o}n Operativa, Universidad Miguel Hern\'{a}%
ndez, 03202 Elche, Spain}
\email{rdale@umh.es}
\author[Jos\'{e} M. Amig\'{o}]{Jos\'{e} M. Amig\'{o}}
\address{Centro de Investigaci\'{o}n Operativa, Universidad Miguel Hern\'{a}%
ndez, 03202 Elche, Spain}
\email[Corresponding author]{jm.amigo@umh.es}

\title[ Blockchain Metrics ]{Blockchain Metrics and Indicators in Cryptocurrency Trading}
\maketitle

\date{\today}

\begin{abstract}
The objective of this paper is the construction of new indicators that can be useful to operate in the
cryptocurrency market. These indicators are based on public data obtained from the blockchain network,
specifically from the nodes that make up Bitcoin mining. Therefore, our analysis is unique to that network.
The results obtained with numerical simulations of algorithmic trading and prediction via statistical models and
Machine Learning demonstrate the importance of variables such as the hash rate, the difficulty of mining or the
cost per transaction when it comes to trade Bitcoin assets or predict the direction of price. Variables obtained
from the blockchain network will be called here blockchain metrics. The corresponding indicators (inspired by
the \textquotedblleft Hash Ribbon \textquotedblright) perform well in locating buy signals. From our results, we conclude that such blockchain
indicators allow obtaining information with a statistical advantage in the highly volatile cryptocurrency
market.
\end{abstract}

\tableofcontents

\maketitle

\section{Introduction{\protect\normalsize \ }}

\label{s:intro}

Since the foundation of the first stock exchanges in Amsterdam and London,
financial analysts and professional traders have been interested in the
development of mathematical indicators that help to anticipate the direction
of the price and thus open positions in the right direction. These
mathematical indicators can be divided into three types: patterns,
oscillators and financial indicators (or just indicators from here on).
Next, we briefly elaborate on these three concepts.

\textit{Patterns} are visual representations of the price history. Usually, the price fluctuates in a range between support (upper endpoint) and resistance (lower endpoint) and forms there a pattern, most of which have fanciful names such as double or triple top/bottom, head and shoulders, ascending or descending triangle, etc. \cite{edwards2018technical}. Eventually, the price leaves that range, breaking the support or resistance in what is known as a breakout. Breakouts are often traded because statistically prices move in the direction of the breakout. In algorithmic trading, recurrent neural networks or convolutional neural networks are used to identify patterns. As a general rule, to trade in view of a pattern it is necessary to combine it with other indicators and different time-frames \cite{krishnan2009impact}.

\textit{Oscillators} are graphs obtained by mathematical calculations that generally oscillate between two values, hence their name. For example, the Relative Strength Index (RSI) is an oscillator that varies between 0 and 100. Other examples include the Moving Average Convergence Divergence (MACD) and the Average Directional Index (ADX). Oscillators use to be placed below the price graphs \cite{vaiz2016study}.

Finally, \textit{Indicators} are metrics in form of lines, commonly obtained by moving averages, that accompany the price evolution. These lines are used to identify trends, analyze the deviation of the actual price in a trend, or simply as supports and resistances. Classical indicators in this category include the Bollinger bands \cite{bollinger1992using} and the Ichimoku cloud \cite{patel2010trading}.

Needless to say, the purpose of moving averages in the indicators is to
filter noise and volatility in the prices. In the 19th and 20th centuries,
authors such as C.H. Dow and R.N. Elliott began using moving averages to
price values and identify primary and secondary trends \cite{sheimo2020cashing}. Nowadays, moving averages are one of
the main technical tools employed by professional traders to speculate on
the stock market.

The possibility of anticipating prices or simply price directions in
financial markets by the analysis of charts, patterns, oscillators and
financial indicators has given rise to an interesting debate about the real
efficiency of the markets. On one hand, the Efficient Market Hypothesis
(EMH) \cite{fama1970efficient} maintains that the
prices of financial assets reflect all the information available in the
market. This hypothesis is challenged by the Adaptive Market Hypothesis
(AMH) \cite{lo2004adaptive}, which maintains that
financial markets are not completely efficient, so prices can drift from
their \textquotedblleft fair\textquotedblright\ value. Hence, EMH and AMH
offer different approaches to understanding the efficiency of financial
markets and have significant implications for how investors perceive and
participate in the markets. Since this is a paper on financial indicators in
cryptocurrency trading, we would like to be more specific on those
implications and take a stance on that debate.

EMH takes three forms: weak, semi-strong and strong. According to the
semi-strong form, which is the most relevant to this paper, current prices
incorporate all public information, including past prices and any other
publicly accessible information. Therefore, market prices are unpredictable,
since it is not possible to obtain returns consistently higher than the
average calculated on past or public information. In particular, investors
cannot beat the market by analyzing technical or fundamental data.

Unlike the EMH, the AMH recognizes the possibility of short-term arbitrage
opportunities due to temporary price inefficiencies. These inefficiencies
can arise due to factors such as investor behavior, market psychology, or
the availability of new information that is not fully reflected in prices.
Moreover, the adaptation of prices to the information available can be slow
and is often subject to temporary fluctuations caused by buying or
overselling due to fear or greed.

In this paper, we propose the use of machine learning algorithms in Bitcoin
price prediction via data from the blockchain network. Incorporating
blockchain network data into cryptocurrency price prediction algorithms
shows how emerging technologies can influence the debate between the EMH and
the AMH. On one hand, the blockchain network provides a high degree of
transparency, since anyone can access and verify transactions. This may
support the EMH since information becomes a more widely available and
actionable resource. However, a scenario with significant delays in
incorporating the blockchain information into cryptocurrency prices is more
favorable to the AMH. Our results shows that the blockchain network allows
obtaining information with a statistical advantage in the highly volatile
cryptocurrency market, which position us in favor of the AMH.

Since the advent of the blockchain network \cite{sarmah2018understanding} and Bitcoin \cite{nakamoto2008bitcoin} in this century, the amount of blockchain-based
data that can be correlated with the Bitcoin price has increased
dramatically. In a nutshell, the blockchain network is a network of
computers connected through protocols that form a decentralized accounting
technology. Indeed, this network records any digital asset or transaction
without the need for a centralized system to create and manage the assets
and transactions. Before the creation of the blockchain network, the
implementation of algorithmic trading was mainly based on historical price
and volume data. With the inception of the blockchain network, additional
data has been made available containing information about the network
itself. These datasets, sometimes called \textit{blockchain metrics}, qualify in principle for Bitcoin price
predictors. On the other hand, blockchain metrics are even more noisy and
volatile than the prices themselves, so smoothing the raw data is more
necessary than ever before beginning any analysis. In fact, data smoothing
are built into any \textit{blockchain indicator},
i.e., an indicator that operates with blockchain metrics.

One of the prototypes of a blockchain indicator is the so-called \textit{Hash Ribbon} \cite{charles_caprile_2020}, 
the name referring to the hashing rate in the blockchain
network. Specifically, the Hash Ribbon consists of two moving averages with
long and short periods whose crossings signalize the long and short
positions, meaning that the holder of the position will profit if the value
of the asset rises or falls, respectively. Inspired by the Hash Ribbon, in
this paper we extend the above methodology to include other blockchain
metrics which, as we will show, help improve profitability in algorithmic
trading systems for crypto assets. Examples include the hash rate as well as
the difficulty of mining, the cost per transaction and many more. The result
are new indicators called \textit{blockchain
ribbons}. Although the focus in this paper is on the Bitcoin, our approach
can be applied to other cryptocurrencies, too.

To explore the potential of blockchain metrics and ribbons, we run two
numerical tests in this paper: (i) algorithmic trading with blockchain
ribbons, where the buy and sell signals activated by our \textquotedblleft ribbon approach\textquotedblright are backtested to measure their
performance and profitability; and (ii) prediction, where the blockchain
metrics are used as variables of the Random Forest statistical model as well
as features of a Machine Learning model. The results allow us to identify
the best blockchain metrics, which do not always coincide with those that
best correlate with the Bitcoin price. The blockchain metrics used in the
aforementioned tests are available on the Quandl platform \cite{URLQuandl}, although some of them had to be conveniently
modified to make them amenable to the ribbon technique. The Python codes
used in the numerical simulations are available at \cite{JCKingGithub}.

In conclusion, this is a paper on financial time series analysis whose main
objective is two-fold: (i) to highlight the importance of the blockchain
metrics when it comes to cryptocurrency trading and (ii) show with numerical
simulations the satisfactory performance of the corresponding blockchain
indicators both in algorithmic trading and prediction. Although the emphasis
is on mathematical issues, the financial terms are conveniently explained
and illustrated as needed, to make the exposition self-contained. As a
result, this paper is also suitable for the growing number of data analysts
interested in the new features introduced by cryptocurrencies.

The rest of this paper is organized as follows. In Section 2, we begin the
technical exposition with some basic notions of the Bitcoin technology that
will suffice as background for the following sections. In particular, the
notion of hash will allow us to introduce in Section 3 the Hash Ribbon,
which is the prototype of an indicator based on blockchain metrics to obtain
long and short signals. Sections 4 and 5 are devoted to briefly describe and
test the new indicators, i.e., the results of applying the ribbon approach
to the blockchain metrics, in some cases after further processing of the
metrics (Section 5). The tests consist of numerical simulations of long and
short trading. In Chapter 6, the blockchain metrics are also tested with
predictive models, namely, Random Forest and long short-term memory (LTSM)
networks. To this end, historical values of blockchain metrics are fed into
both models and their predictions of the closing price of Bitcoin are compared
with a prediction horizon of 10 days. The last two sections present a
discussion of the results (Section 7) and the conclusions (Section 8).

\section{Bitcoin technology}

{\normalsize \label{s:history.2} }

In a nutshell, Bitcoin is an electronic money system for peer-to-peer payments. Bitcoin is the currency itself, analogous to the Dollar or the Euro, and the blockchain network is the network of computers (nodes) that supports the assets and transfers, as the conventional banking network does. However, there are major differences between both networks. 

Thus, in a conventional banking network, the user has a private key that allows him/her to order a transfer, while the issuing entity is responsible for making said transfer. In this case, the control of the money movements is held by the bank and it marks the costs of the transfer. On the other hand, to make transfers in the Bitcoin network, the user also needs a private key to order a transfer, but now the request must be approved by the majority of the nodes that make up the network. Therefore, the control is no longer central, and the transaction costs, that are determined by an algorithm (the Bitcoin algorithm), are randomly paid to some of the nodes that have validated the transaction. Once validated, the transaction becomes effective and registered in all the databases that make up the blockchain ledger of the Bitcoin network. The more nodes, the more secure the network is against cyberattacks and, consequently, the more valuable the assets.

The Bitcoin algorithm was designed to create approximately 21 million Bitcoins. Bitcoins are the rewards to the mining nodes for decoding blocks, each block containing a SHA-256 cryptographic hash of the previous block up to the first block (which justifies the name of blockchain). A \textit{hash function} is a computationally efficient function mapping binary strings of arbitrary length to binary strings of some fixed length (256 bits in the case of the algorithm SHA-256), called \textit{hash-values} or simply \textit{hashes} \cite{menezes1997}. At the beginning of this technology, i.e., in the period 2009-2011, mining was very simple and Bitcoins had zero value. At that time it was possible to mine with any computer infrastructure. But as time passed and Bitcoin became more valuable, the difficulty of mining increased considerably. Currently mining is only possible with large computational infrastructures. For more details, see e.g. \cite{mackenzie2019pick}.

Finally, let us point out that there are three ways to acquire Bitcoins:

\begin{itemize}
\item {By buying Bitcoins directly from an exchange.  }
\end{itemize}

\begin{itemize}
\item {By mining a block and receiving the reward in Bitcoins. To this end, it is necessary to add a node to the blockchain network.}

\item {By validating a transaction, which also requires to add a new node to the blockchain network.}
\end{itemize}

In summary and for the purposes of this paper, the Bitcoin technology is implemented by a computer network called the blockchain network. The value of the Bitcoin is created by mining blocks at the nodes of the network, which consists of retriving a text from its SHA-256 hash. We will use below information related to the blockchain.


\section{Methods}

\label{s:methods}

We call \textit{indicator} the result of processing historical Bitcoin price data or other related datasets to identify trends and signals in the market. Indicators are represented in graphs of the price against time by lines. In our case, the datasets include the Bitcoin daily closing price since 2012 (when its trading began), along with data extracted from the blockchain network. The final goal of the indicators introduced in the present work is to trade in the Bitcoin market. The methodology that we followed consisted of the following steps. 

After surveying data from the Bitcoin blockchain network, we selected datasets obtained through Quandl; we chose all the datasets available on that platform belonging to the mining of the Bitcoin network nodes. Quandl is a platform that collects daily data from the blockchain network of Bitcoin and exploits it via visualizations; Quandl has recently been acquired by the Nasdaq platform. These datasets are the \textit{blockchain metrics} that we will consider in this paper (Section \ref{s:methods.1}).

The blockchain metrics were then processed with quantitative tools such as nonlinear coefficients of functional dependency and moving averages to extract information that correlates with the Bitcoin price. We also tested the ribbon technique, which resorts to moving averages of different periods, as explained for the hash ribbon in Section \ref{s:history.3.1}. To this end, four blockchain metrics had to be refined via linear regression or derivatives (Section \ref{s:applications}). The result are the new indicators or \textquotedblleft blockchain ribbons\textquotedblright\ that we propose in Sections \ref{s:blockchain ribbons.3.4} and \ref{s:applications}.

Finally, we did two types of numerical simulations. First, simulations of buying and selling with the new ribbons to numerically check their quality, performance, advantages and disadvantages in algorithmic trading (Section \ref{s:methods.4}). Second, predictions with recurrent neural networks and Random Forest to test how well the blockchain metrics anticipate the price direction (Section \ref{s:predictions}). For both simulations we use all blockchain metrics to check whether nonlinear correlation alone can capture their full potential in trading and prediction.   

We begin next with the blockchain metrics and the functional dependency of the Bitcoin price on them.


\subsection{Blockchain metrics}

\label{s:methods.1}

In this work we use all datasets available and published on the Quandul Platform [\cite{URLQuandl}]. Records have been accessed using the public Rest API. The datasets, alphabetically ordered by their acronyms, are the following: 

\noindent (1) Median Transaction Confirmation Time (ATRCT); (2) Average Block Size (AVBLS); (3) Api Block-Chain Size (BLCHS); (4) Cost Per Transaction (CPTRA) ; (5) Difficulty (DIFF); (6) Estimated Transaction Volume (ETRAV); (7) Estimated Transaction Volume USD (ETRVU); (8) Hash Rate (HRATE); (9) Miners Revenue (MIREV); (10) Market Price USD (MKPRU); (11) Market Capitalizacion (MKTCP); (12) My Wallet Number of Users (MWNUS); (13) Addresses Used (NADDU); (14) Number of Transactions (NTRAN); (15) Total Number of Transactions (NTRAT); (16) Number of Transaction per Block (NTRBL); (17) Number of Transactions Excluding Popular Addresses (NTREP); (18) Total Output Volume (TOUTV); (19) Total Transaction Fees (TRFEE); (20) Total Transaction Fees USD (TRFUS); (21) USD Exchange Trade Volume (TRVOU). 

Note that the market price of the Bitcoin (actually the daily closing price) is the 10th dataset (MKPRU). Some of these datasets have been investigated in \cite{detzel2021learning}. 

When we speak of blockchain metrics hereafter, we mean the above 21 datasets. These datasets contain daily values of the corresponding blockchain metric from January 2d, 2009 to October 3rd, 2022. However, the time series that we analyzed start on August 16th, 2010 ($t=1$) because the Bitcoin price was zero before that date, with only one exception: the time series of the metric ATRCT starts on December 2nd, 2011 because before its values were zero. Therefore, the time series ATRCT has a length of 3,958 records, while the other 20 time series have 4,432 records. Figures \ref{figure2} and  \ref{figure4}-\ref{figure9} show smoothed curves (simple moving averages) of a few blockchain metrics.



\subsection{Study of functional dependencies}

\label{s:methods.2}

To identify those datasets that may have the most predictive power for the
Bitcoin price we computed the coefficient $\xi _{n}(X,Y)$ of Chatterjee \cite
{Chatterjee2021}, where $X$, $Y$ are random variables and $n\geq 2$.  

The coefficient $\xi _{n}(X,Y)$ is an asymptotic estimator of functional
dependence. Specifically, $\lim_{n\rightarrow \infty }\xi _{n}(X,Y)=0$ if
and only if $X$ and $Y$ are independent, and $\lim_{n\rightarrow \infty }\xi
_{n}(X,Y)=1$ if and only if $Y$ is almost surely equal to a measurable
function of $X$. The asymptotic estimator $\xi _{n}(X,Y)$ is computed from $%
n\geq 2$ i.i.d. realizations $(x_{1},y_{1})$, ..., $(x_{n},y_{n})$ of $(X,Y)$%
. In our analysis, $y_{t}$, $1\leq t\leq n$, is the $t$-th daily closing
price of the Bitcoin (from the dataset MKPRU), while $x_{t}$ is the
corresponding entry in any of the other 21 datasets. Table \ref {table1} shows the results obtained for $n$ being the size of the dataset ($4,432$ in most
cases). 

\begin{table}[htbp]
\centering
\begin{tabular}{|>{\centering}p{2cm}|p{3cm}|}
      	\hline
	
	\textbf{Metric}		& 
	\centering  \textbf{$\xi$}  \arraybackslash \\
	\hline
	    MWNUS &\centering 0.986782 \tabularnewline
	    NTRAT &\centering 0.985657 \tabularnewline
	    DIFF  &\centering 0.976248 \tabularnewline
	    MKTCP &\centering 0.973415 \tabularnewline
	    BLCHS &\centering 0.955751 \tabularnewline
	    HRATE &\centering 0.901979 \tabularnewline
	    MIREV &\centering 0.871177 \tabularnewline
	    ETRVU &\centering 0.803517 \tabularnewline
	    TRFUS &\centering 0.754157 \tabularnewline
	    NADDU &\centering 0.740678 \tabularnewline
	    AVBLS &\centering 0.698378 \tabularnewline
	    CPTRA &\centering 0.637157 \tabularnewline
	    NTREP &\centering 0.620065 \tabularnewline
	    NTRAN &\centering 0.612181 \tabularnewline
	    NTRBL &\centering 0.600631 \tabularnewline
	    TRVOU &\centering 0.503343 \tabularnewline
	    ATRCT &\centering 0.339253 \tabularnewline
	    TRFEE &\centering 0.284107 \tabularnewline
	    TOUTV &\centering 0.225585 \tabularnewline
	    ETRAV &\centering 0.124307 \tabularnewline

    \hline
    \end{tabular}
\caption{Coefficient of functional dependency (Chatterjee coefficient of correlation $\xi$) of the Bitcoin price with respect to the other 20 blockchain metrics. The metrics are listed from highest to lowest coefficient. See Section \protect\ref{s:methods.1} for the meanings of the acronyms.} 
\label{table1}
\end{table}

An interesting result of this analysis is that, even though the hash rate (HRATE) is the blockchain metric on which the popular Hash Ribbon indicator was created (see Section \ref{s:history.3.1} below), other metrics have higher Chatterjee coefficients. In particular, the data corresponding to My Wallet Number of Users (MWNUS) is the most (nonlinearly) correlated with the Bitcoin price (MKPRU), followed by NTRAT , DIFF, MKTCP and BLCHS. 

The conclusion is that the Bitcoin price has a good correlation with some blockchain metrics (in the sense of functional dependence on them). Therefore, the blockchain network can provide information with the same or better predictive power than those currently used or published. This conclusion will put to test in Sections \ref{s:methods.4} and \ref{s:predictions}, where we will perform some numerical simulations of algorithmic trading and prediction using all the blockchain metrics.

\subsection{Hash ribbon: the prototype of blockchain ribbons}

\label{s:history.3.1}

Similar to the hash ribbon already in use, the \textit{blockchain ribbons} consist of two moving averages of long and short periods whose crossings determine the long and short signals, i.e., when to buy and when to sell, respectively. These are the new indicators that we are going to use in Section \ref{s:methods.4} to determine the long and short signals in the Bitcoin trading. Therefore, we will revisit the hash ribbon next.

\begin{figure}[th]
\centering
\includegraphics[width=130mm]{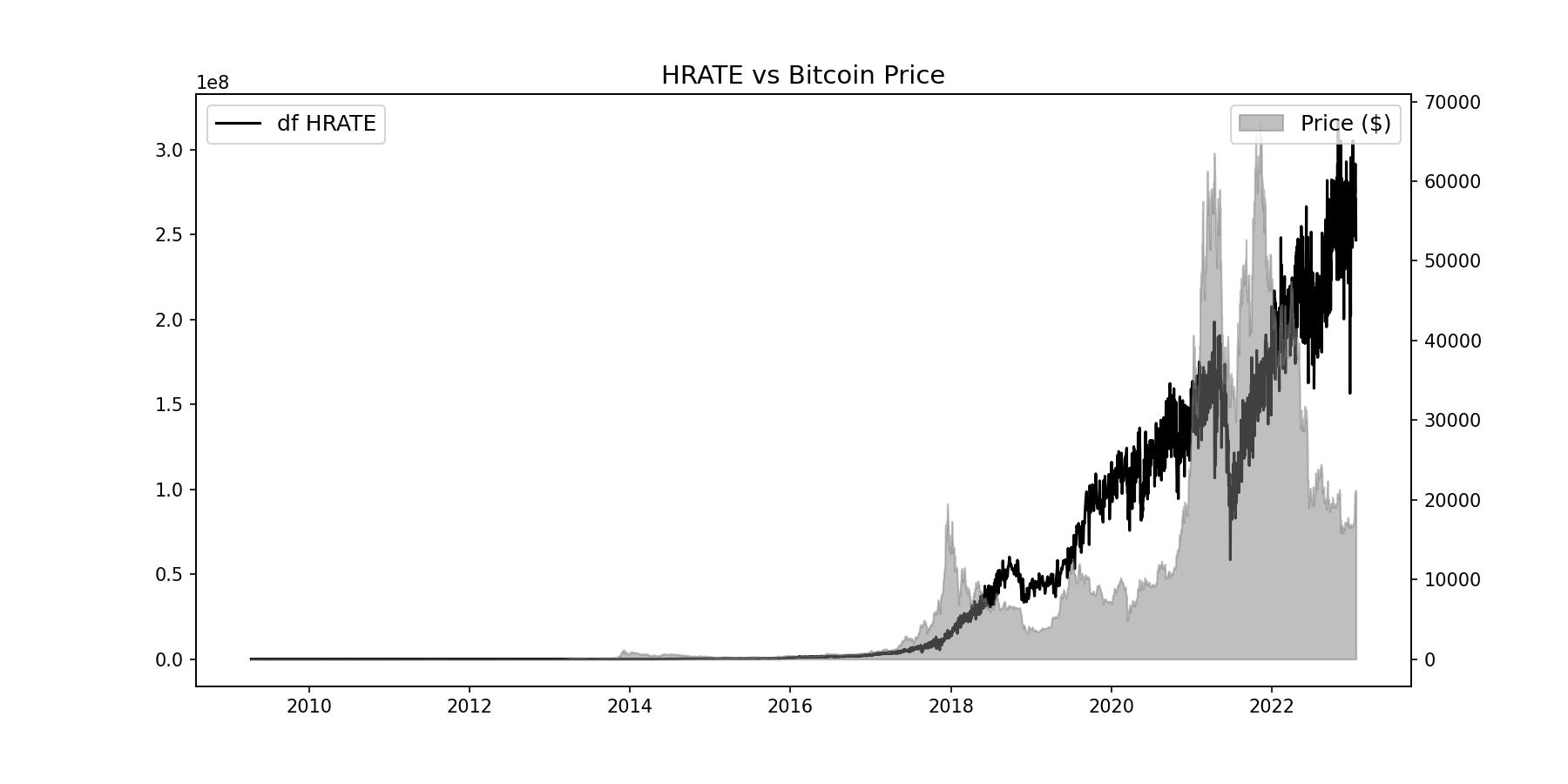}
\caption{Hash rate vs Bitcoin price. The continuous line is the hash rate and the gray area represents the Bitcoin price, see insets.}
\label{figure1}
\end{figure}

\begin{figure}[th]
\centering
\includegraphics[width=130mm]{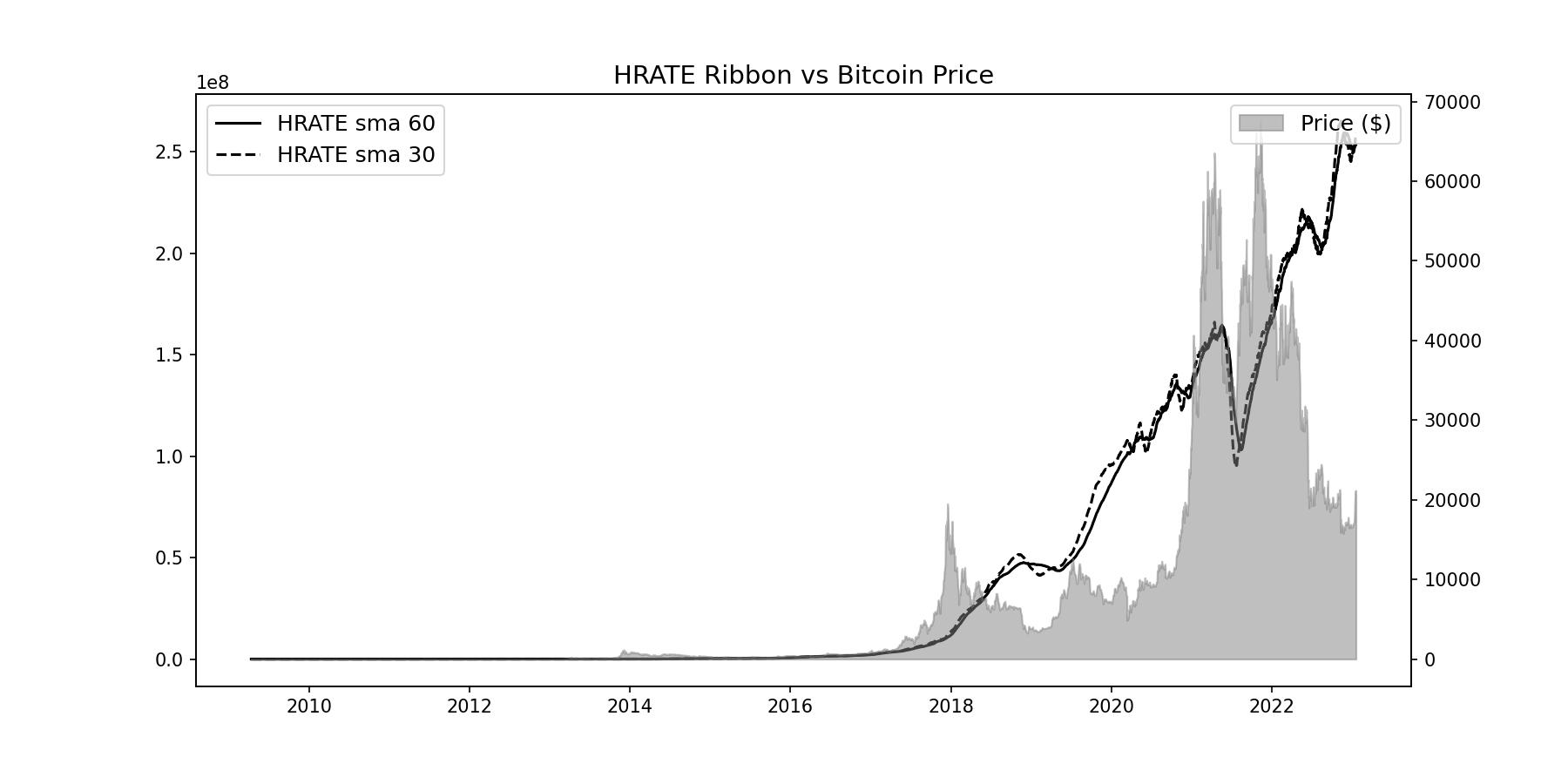}
\caption{Hash ribbon vs Bitcoin price. The long ribbon line (SMA-60) is the continuous one, and the short ribbon line (SMA-30) is the dotted one, see insets.}
\label{figure2}
\end{figure}

The hash ribbon (HR) is an indicator created by the trader and analyst Charles Edwards, who published its code in the TradingView portal in 2019 \cite{charles_caprile_2020}. Since this indicator is very recent, there is no official information about it. However, it follows from the code that the raw data comes from the Quandl platform and the data collected is the hash rate of the blockchain network.

The hash rate \cite{fantazzini2020does} indicates how fast the miners' machines are performing calculations to decode the blocks. The hash rate is measured by the number of calculations that the blockchain network performs per second; its units are Megahash per second (MH/s), Gigahash per second (GH/s) and Terahash per second (TH/s). Thus, a hash rate of 1 TH/s means that the blockchain network is performing $10^{12}$ calculations per second. As an additional remark, the more computers compete to obtain rewards and validate transactions, the more secure the blockchain network becomes. Therefore, the higher the hash rate, the less vulnerable the network is \cite{dev2014bitcoin}.

As shown in Figure \ref {figure1}, the raw hash rate data is too noisy to be useful for identifying patterns. The traditional method for smoothing out noisy data is to take moving averages (MA). The hash ribbon consists of two such averages: $HR_{\min }(t)$, which is a simple moving average of the last 30 days (the short period MA), and $HR_{\max }(t)$, which is a simple moving average of the last 60 days (the long period MA), i.e.,

\begin{equation}
HR_{\min }(t)=\frac{1}{30}\sum_{i=1}^{30}HR(t-i),\;\;HR_{\max }(t)=\frac{1}{%
60}\sum_{i=1}^{60}HR(t-i),  \label{HRmin}
\end{equation}%

\noindent where $HR(i)$ is the average hash rate on day $i$. The result is shown in Figure \ref{figure2}. The crossings of $HR_{\min }(t)$ and $HR_{\max }(t)$ are used as long signals (when $HR_{\max }(t)$ crosses from below) and short signals (when $HR_{\max }(t)$ crosses from above).

\begin{figure}[th]
\centering
\includegraphics[width=120mm]{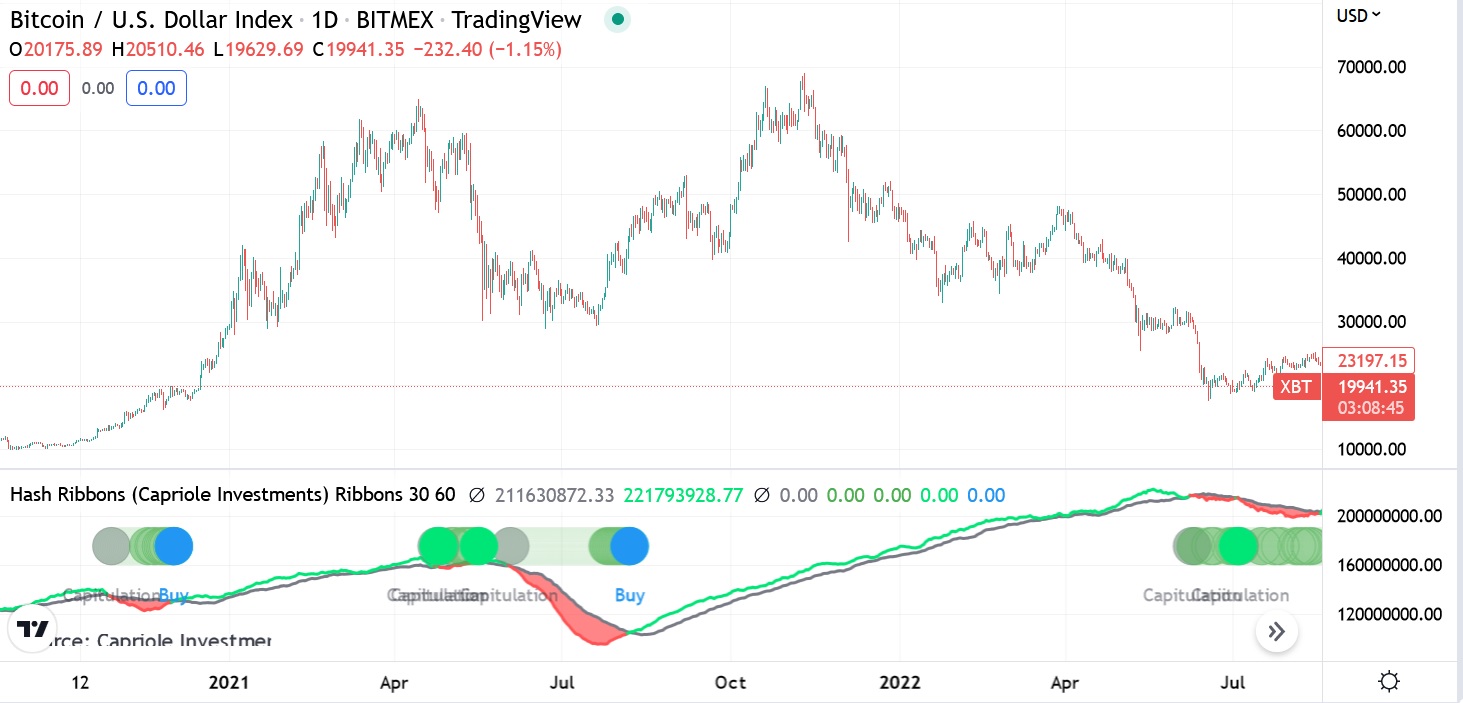} 
\caption{Actual visualization of the hash ribbon.}
\label{figure3}
\end{figure}

Usually, the moving averages $HR_{\max }(t)$ and $HR_{\min }(t)$ are displayed below the price chart, see Figure \ref {figure3}, so that the trader can conveniently make decisions; they are called the long and short \textit{ribbon lines}, respectively. It can be checked that the signals are quite accurate, particularly the long ones.


\subsection{Blockchain ribbons}

\label{s:blockchain ribbons.3.4}

Blockchain ribbons are generalizations of the hash ribbon to other blockchain metrics that are amenable to the ribbon technique. They comprise all the blockchain metrics except the MWNUS, NTRAT and BLCHS metrics because these are monotonic functions of the Bitcoin price, as we will show shortly. For illustration and further reference, we will also describe the CPTRA, DIFF and MIREV ribbon. All these metrics, along with the HR ribbon, have nonlinear correlation coefficients above 0.6 (see Table \ref{table1}); in particular, the MWNUS and NTRAT metrics have the highest coefficients. As with the HR ribbon, long lines (SMA-60) will be depicted as continuous lines, while short lines (SMA-30) as dotted lines in the figures.

\begin{figure}[th]
\centering
\includegraphics[width=120mm]{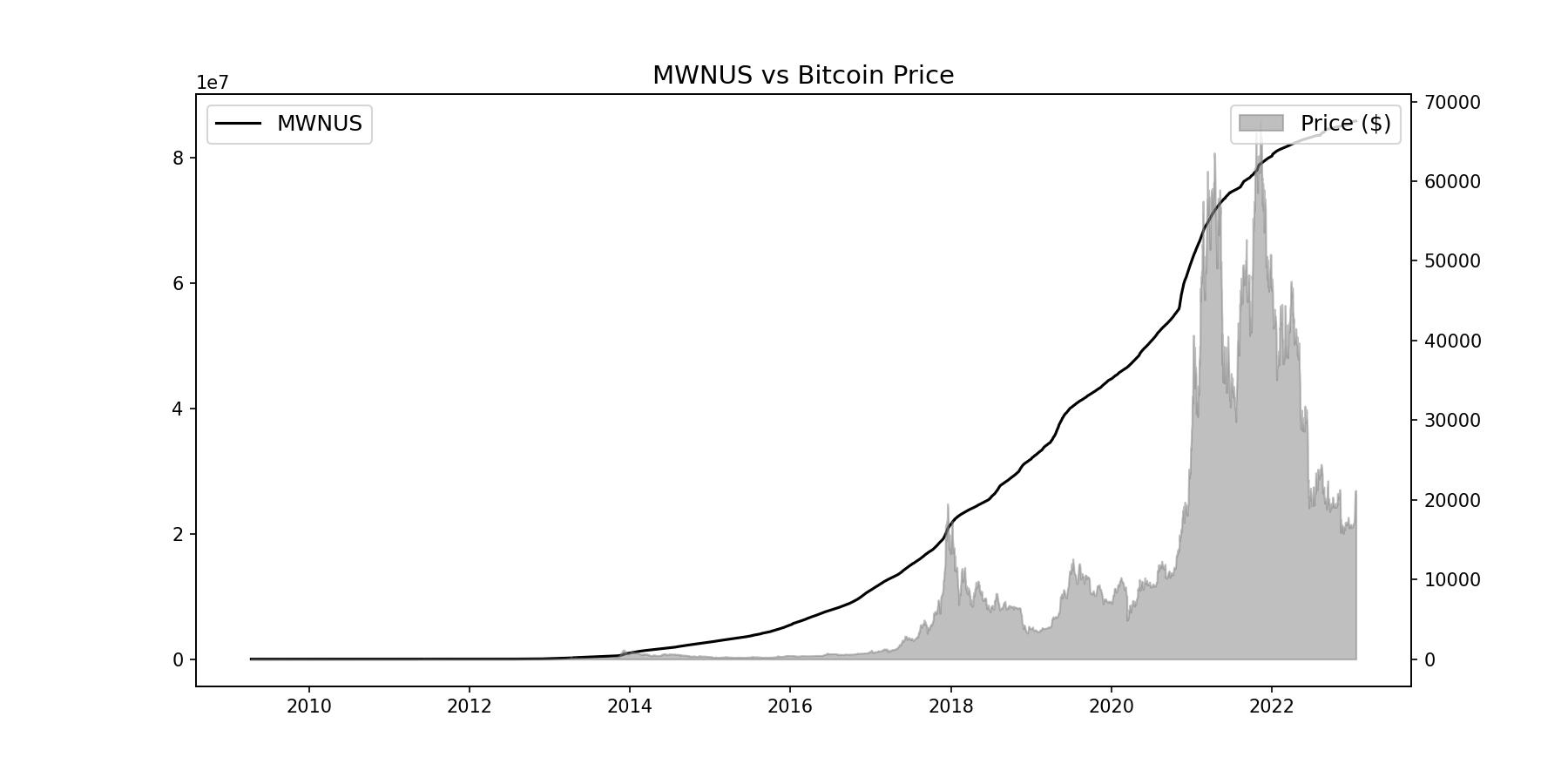}
\caption{Bitcoin My Wallet Number of users vs Bitcoin Price. }
\label{figure4}
\end{figure}

\begin{figure}[th]
\centering
\includegraphics[width=120mm]{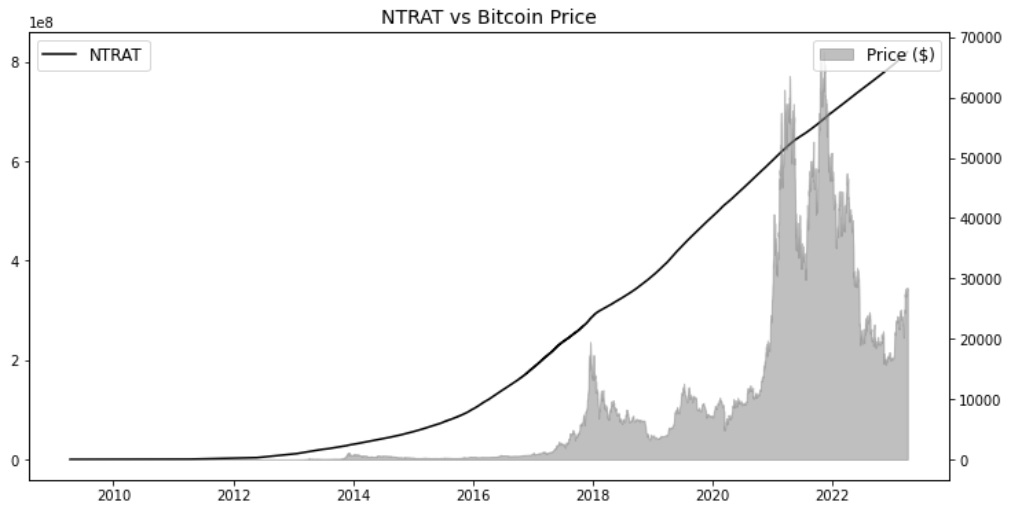}
\caption{ Total Number of Transactions vs Bitcoin Price.}
\label{figure5}
\end{figure}

\begin{figure}[th]
\centering
\includegraphics[width=120mm]{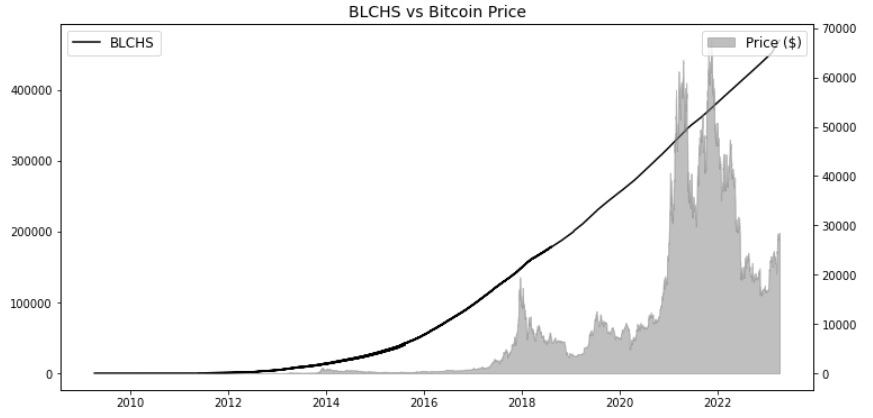}
\caption{Api Block-Chain Size vs Bitcoin Price.}
\label{figure6}
\end{figure}

Thus, first of all we consider the metric My Wallet Number of Users (MWNUS), which counts the number of new users entering the market by creating a wallet. Despite having the highest Chatterjee correlation coefficient (see Table \ref{table1}) and the fourth highest Pearson coefficient (not shown), MWNUS does not rise or fall with the price of the Bitcoin but, instead, increases monotonically with time, see Figure \ref{figure4}. Therefore,

\begin{eqnarray}
MWNUS_{\max }(t) &=&\frac{1}{60}\sum_{i=1}^{60}MWNUS(t-i)  \nonumber \\
&=&\frac{1}{60}\left( 30MWNUS_{\min }(t)+30MWNUS_{\min }(t-30)\right)  
\nonumber \\
&=&\frac{1}{2}\left( MWNUS_{\min }(t)+MWNUS_{\min }(t-30)\right) 
\label{monotonicity} \\
&\leq &\frac{1}{2}\left( MWNUS_{\min }(t)+MWNUS_{\min }(t)\right)   \nonumber
\\
&=&MWNUS_{\min }(t)  \nonumber
\end{eqnarray}

\noindent where the monotonicity of the data was used on the last but one line of Equation \ref{monotonicity}. This shows that the MWNUS ribbon lines cannot cross. 

\begin{figure}[th]
\centering
\includegraphics[width=130mm]{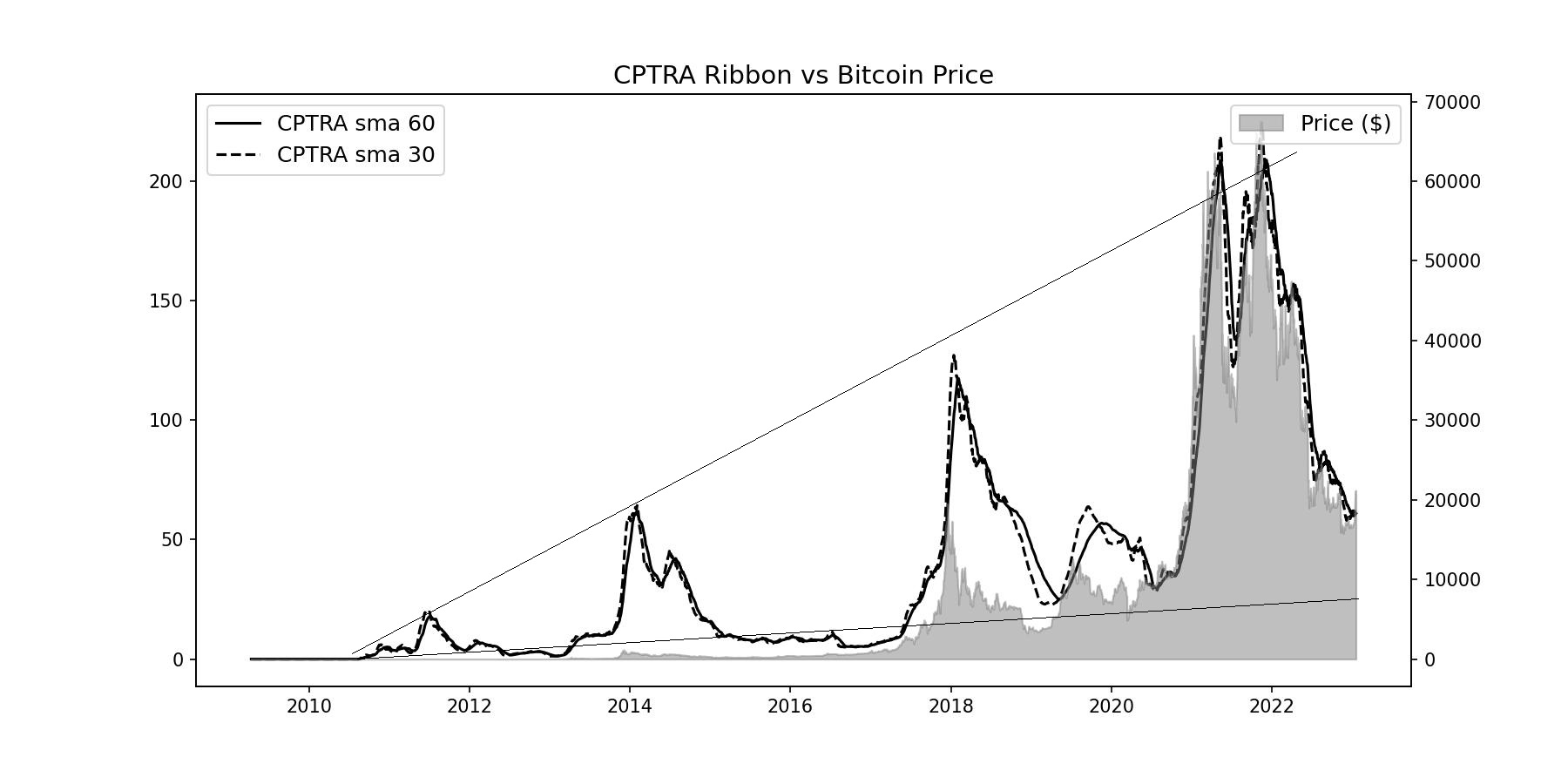}
\caption{Bitcoin Cost Per Transaction vs Bitcoin Price. The relevance of the lines drawn through certain maxima and minima is addressed in the text and in Section \ref{s:applications.1}.}
\label{figure7}
\end{figure}

\begin{figure}[th]
\centering
\includegraphics[width=130mm]{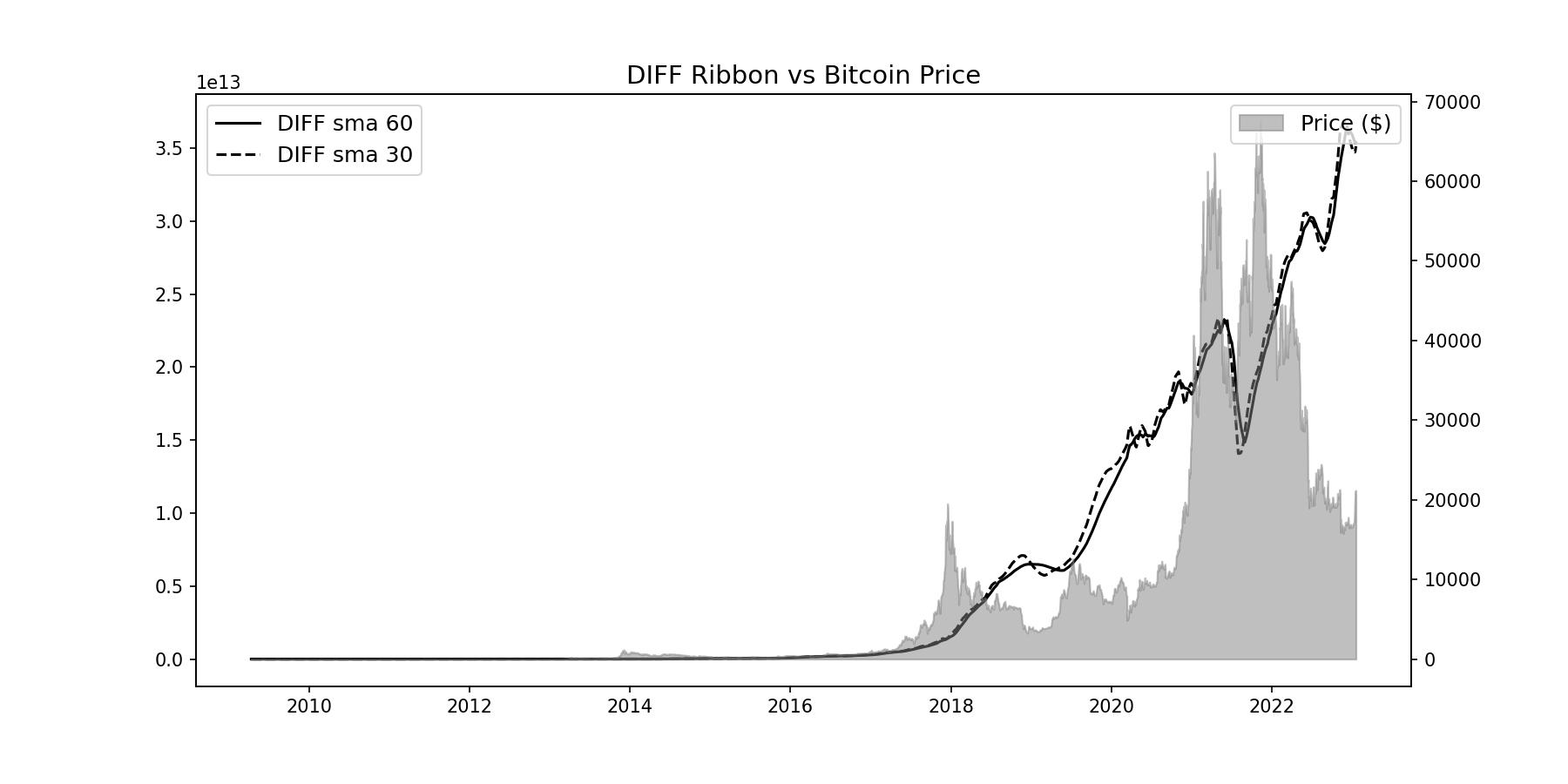}
\caption{ Bitcoin Difficulty vs Bitcoin Price}
\label{figure8}
\end{figure}

\begin{figure}[th]
\centering
\includegraphics[width=130mm]{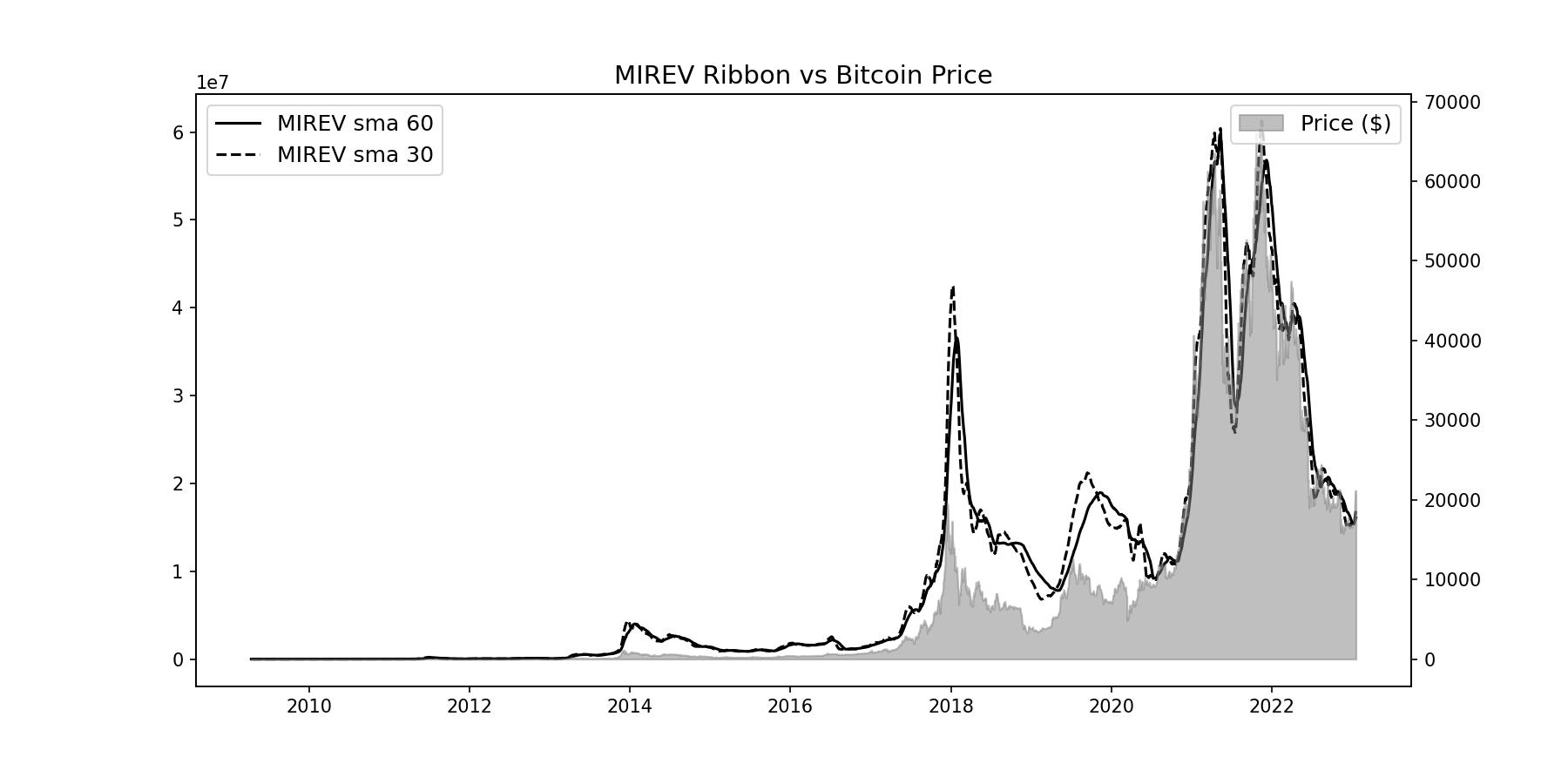}
\caption{Bitcoin Miners Revenue vs Bitcoin Price.}
\label{figure9}
\end{figure}

The same happens with the metric Total Number of Transactions (NTRAT), see Figure \ref{figure5}, and the Blockchain Size (BLCHS), see Figure \ref{figure6}. Therefore, this metric cannot be promoted to a blockchain ribbon either, while all the other blockchain metrics considered in this paper can; a few examples follow.

An example of a blockchain metric that is amenable to the ribbon technique is the Bitcoin Cost Per Transaction. The CPTRA ribbon, Figure \ref{figure7}, depends on the price, the block size, and the amount of transactions or traffic in the network. In addition, the straight line drawn through ever higher local maxima since 2011 has been a resistance zone where the price suffered a strong rejection. In fact, there were five times in which the CPTRA metric approached this zone and in all of them the price was rejected. Therefore, the CPTRA ribbon can be used to operate through long and short signals or to move away from areas where the price may have big drops.

As a second example, the Difficulty (DIFF) metric, Figure \ref{figure8}, is a measure of how difficult it is to find or mine a Bitcoin block. Technically, this amounts to finding a hash determined by the algorithm at that time. A high network difficulty means that it takes more computational cost to mine the same amount of blocks, which makes the network more secure against cyberattacks but more expensive for miners.

Finally, the Bitcoin Miners Revenue (MIREV) ribbon, Figure \ref{figure9}, refers to the proportion in economic compensation that a miner receives for decoding a block, as well as the total value of rewards and transaction fees paid to them. It is a very significant piece of information to check the profitability of mining farms. In this case the long and short signals are also apparently correct as shown in Figure \ref{figure9}. Furthermore, it is possible to identify a straight line by joining the 2014 and 2018 all-year highs of the long ribbon line, which in the year 2022 was a zone of resistance where the price and the metric had a strong rejection.


\section{Performance of the blockchain metrics using the ribbon technique}

\label{s:methods.4}

In this section we test the performance of the blockchain ribbons (Section \ref{s:blockchain ribbons.3.4}) as indicators of long and short positions, i.e., its suitability for algorithmic trading. We will also discuss the results. Similar to the HR ribbon (Section \ref{s:history.3.1}), the way to operate in the market is as follows. If the long line crosses the short line from below, then the market is bullish and the recommended action is to buy; we speak of a long position and a long trade or operation. If, otherwise, the long line crosses the short line from above, then the market is bearish and the recommended action is to sell; we speak of a short position and a short trade or operation.


\subsection{Numerical simulations of long and short trading}

\label{s:long and short.4.1}

We are going to simulate trading operations using the blockchain ribbons. Therefore, a crossing of the ribbon lines signalizes an opportunity to enter the market in a long or short position, while the next crossing is a signal to exit and/or (re)enter in the opposite position. The aim of our simulations is to quantify the predictive ability of the blockchain metrics by the potential profit margins. The Python codes used in the numerical simulations are available at \cite{JCKingGithub}.

Simulations begin on August 16th, 2010, the day Bitcoin goes from \$0 to \$0.0769, and ends on December 31, 2022. At that time, and after much volatility, the value is approximately \$20,000. Each of the blockchain ribbons has been considered separately, analyzing the buy and sell signals separately, and the performance obtained for each of them has been measured.

For each operation, there can be four possible results.

A. \textit{Winning trade}, which means that the anticipated price direction was correct and the trade was closed with a positive profit margin. For example, a long trade that opens at \$20,000 and closes at \$21,000.

B. \textit{Losing trade}, which means that the anticipated price direction was not correct and the trade was closed with a negative profit margin. For example, a short trade that opens at \$20,000 and closes at \$21,000.

C. \textit{Stop Loss Activated} is the worst scenario. It occurs when the market moves abruptly in the opposite direction than anticipated and the price reaches the stop loss threshold. For example, a short trade is opened at \$20,000 and the market reaches a threshold of \$26,000. 

D. \textit{Target Achieved} is the most desired scenario. It occurs when a high enough profit target has been reached to close the trade and the trader or algorithm waits for the next signal.

In our simulations, we have calculated the following \textit{performance indicators} for each long and short operation.

1. \textit{Maximum Price} is the maximum value of the Bitcoin price (in percentage) between two consecutive crossings of the ribbon lines.

2. \textit{Minimum Price} is the minimum value of the Bitcoin price (in percentage) between two consecutive crossings of the ribbon lines.

3. \textit{Trade Profit} is the difference in the Bitcoin price between the exit price and the entry price, depending on whether the initial position is long or short, namely:
\begin{equation}
\text{Long Profit = Exit Price - Entry Price}  \label{Long Profit}
\end{equation}

and%
\begin{equation}
\text{Short Profit = Entry Price - Exit Price}  \label{Short Profit}
\end{equation}

In addition, we have calculated the following quantities:

4. \textit{Number of Trades (n)}.

5. \textit{Number of winning trades (w)}. Given two blockchain ribbons that offer a similar and positive profit margin, the total profit margin is greater for the ribbon that signals more winning trades.

6. \textit{Number of losing trades (l)}.

7. \textit{Percentage of Winning Trades (WT)}, i.e.,
\begin{equation}
WT\,(\%)=100\;\frac{w-l}{n}.
\end{equation}%

8. \textit{Average of the maximum Bitcoin price} (in percentage) in the time from the entry signal to the exit signal of each operation.

9. \textit{Average of the minimum Bitcoin price} (in percentage) in the time from the entry signal to the exit signal of each operation.


Although this paper is not about trading strategies, a one-to-one (1:1) trading strategy has been implemented in the numerical simulations, in which the stop loss threshold and the target are equal to 30\%. This is a strategy in which you assume a statistical advantage and define the same risk as the maximum target value at which each trade will be closed. In our case, if a trade is opened in which the maximum revaluation of the asset is expected to be 30\%, the stop loss must also be 30\%, so that if the market turns around, the maximum loss will be exactly the same as the maximum expected value. In other words, each signal is traded expecting the value to increase by at least $30\%$. This being the case, the following performance indicators has also been calculated.

10. Percentage of times the $30\%$ \textit{target threshold} was \textit{reached}.

11. Percentage of times the $30\%$ \textit{stop loss} was \textit{activated}.

12. \textit{Strategy Balance (SB)}, i.e.,

\begin{equation}
SB\,(\%)=100\;\frac{t-s}{n},  \label{SB}
\end{equation}%
where $t$ is the number of trades with target achieved, $s$ is the number of trades with stop loss activated, and $n$ is the total number of trades.

13. \textit{Total Profit} of the strategy (STP) is defined as
\begin{equation}
\text{STP (\%) = \textit{balance achieved} (BA) + \textit{balance non-achieved}
(BN) - \textit{total fees} (TF).}  \label{STP}
\end{equation}%
Here%
\begin{equation}
BA=\frac{t-s}{t+s} \; Target \; (\%),
\end{equation}%
where \textit{Target} is the maximum value expected (30\%) and $t$ and $s$ are as above,
\begin{equation}
BN=\frac{1}{w}\sum_{i=1}^{w}PT({i}) \; (\%),
\end{equation}%
where $PT({i})$ is the Profit Trade (Equations (\ref{Long Profit})-(\ref {Short Profit})) in the $i$th operation and $w$ is the number of trades without stop loss or target achieved, and finally

\begin{equation}
TF=Trade\,Fees\ast n  \; (\%),
\end{equation}
where, as before, $n$ is the number of trades. Trading fees are around $1\%$
in the cryptocurrency market.

The results for the above performance indicators (columns) and the currently 18 blockchain ribbons (rows) are shown in Table \ref{table2}, without differentiating the type of operation (long or short). Metrics are ranked according to the Total Profit of the Strategy (last column) in the table. The four metrics NADDU, CPTRA, TRFUS and TRFEE outperform the benchmark MKPRU (Bitcoin price). Comparison with Table \ref{table1} shows that they are not among the most correlated metrics with the Bitcoin price. In other words, the most correlated metrics do not necessarily perform the best in algorithmic trading.  

The mising MWNUS, NTRAT and BLCHS metrics will be built into ribbons in Section 5.

\begin{table}
 	\tiny
  	\renewcommand{\arraystretch}{1.3} 
  
  	\resizebox{1\textwidth}{!} {

	\begin{tabular}{|>{\centering}p{1.3cm}|c|>{\centering}p{1.0cm}|>{\centering}p{1.1cm}|>{\centering}p{1.0cm}|>{\centering}p{1.25cm}|>{\centering}p{1.15cm}|>{\centering}p{1.15cm}|>{\centering}p{1.17cm}|>{\centering}p{1.12cm}|>{\centering}p{1.18cm}|>{\centering}p{1.0cm}|>{\centering}p{1.09cm}|p{1.16cm}|}

      	\hline
	
	\textbf{Metric}			& 
      	\textbf{Trades}			& 
      	\textbf{Winning Trades} 		& 
      	\textbf{Winning Trades (\%) }	& 
      	\textbf{Trade Profit (\%)} 			& 
      	\textbf{Maximum Achieved (\%)}		& 
      	\textbf{Minimum BT Price (\%)}			& 
      	\textbf{Threshold achieved} 			&	
     	\textbf{Stop Loss activated}			&
      	\textbf{Threshold achieved (\%)}		& 
      	\textbf{Stop Loss activated (\%)}		&	
      	\textbf{Strategy Balance}				&	
	\textbf{Trades Total Balance Strategy (\%)}	&
	\centering  \textbf{Strategy Total Profit (\%)}  \arraybackslash \\
	\hline
	 \small   &\small  &\small   &\small  &\small  &\small  &\small  &\small  &\small  &\small  &\small   &\small &\small &\centering \small  \tabularnewline 
	 \small NADDU  &\small 127 &\small 65  &\small 51.18 &\small  58.77 &\small 142.92 &\small   7.49 &\small 25 &\small 9 &\small 19.69 &\small  7.09 &\small 16&\small 12.60 &\centering \small 5819 \tabularnewline 
	 \small CPTRA  &\small 182 &\small 94  &\small 51.65 &\small  28.30 &\small  74.38 &\small  12.90 &\small 48 &\small 16&\small 26.37 &\small  8.79 &\small 32&\small 17.58 &\centering \small 4117 \tabularnewline 
	 \small TRFUS  &\small 117 &\small 61  &\small 52.14 &\small  39.80 &\small 107.35 &\small  10.60 &\small 22 &\small 13&\small 18.80 &\small 11.11 &\small 9 &\small  7.69 &\centering \small 3416 \tabularnewline 
	 \small TRFEE  &\small 122 &\small 61  &\small 50.00 &\small  36.11 &\small  93.92 &\small  17.59 &\small 21 &\small 19&\small 17.21 &\small 15.57 &\small 2 &\small  1.64 &\centering \small 2899 \tabularnewline 
	 \small MKPRU  &\small 77  &\small 36  &\small 46.75 &\small  45.26 &\small 102.84 &\small   9.82 &\small 24 &\small 5 &\small 31.17 &\small  6.49 &\small 19&\small 24.68 &\centering \small 2665 \tabularnewline 
	 \small MKTCP  &\small 71  &\small 32  &\small 45.07 &\small  52.02 &\small 116.88 &\small  10.43 &\small 27 &\small 5 &\small 38.03 &\small  7.04 &\small 22&\small 30.99 &\centering \small 2618 \tabularnewline 
	 \small TRVOU  &\small 139 &\small 71  &\small 51.08 &\small  22.79 &\small  56.39 &\small   9.81 &\small 23 &\small 12&\small 16.55 &\small  8.63 &\small 11&\small  7.91 &\centering \small 2561 \tabularnewline 
	 \small ETRVU  &\small 155 &\small 73  &\small 47.10 &\small  18.76 &\small  49.08 &\small   9.35 &\small 24 &\small 12&\small 15.48 &\small  7.74 &\small 12&\small  7.74 &\centering \small 2437 \tabularnewline 
	 \small MIREV  &\small 93  &\small 46  &\small 49.46 &\small  31.31 &\small  76.56 &\small   9.82 &\small 26 &\small 8 &\small 27.96 &\small  8.60 &\small 18&\small 19.35 &\centering \small 2294 \tabularnewline 
	 \small HRATE  &\small 41  &\small 22  &\small 53.66 &\small  88.95 &\small 319.69 &\small  96.51 &\small 15 &\small 7 &\small 36.59 &\small 17.07 &\small 8 &\small 19.51 &\centering \small 1889 \tabularnewline 
	 \small ETRAV  &\small 227 &\small 122 &\small 53.74 &\small   7.47 &\small  19.71 &\small  11.02 &\small 25 &\small 16&\small 11.01 &\small  7.05 &\small 9 &\small  3.96 &\centering \small 1433 \tabularnewline 
	 \small DIFF   &\small 27  &\small 17  &\small 62.96 &\small 104.20 &\small 441.36 &\small 145.38 &\small 12 &\small 7 &\small 44.44 &\small 25.93 &\small 5 &\small 18.52 &\centering \small  957 \tabularnewline 
	 \small NTREP  &\small 139 &\small 64  &\small 46.04 &\small  10.5  &\small  39.08 &\small  13.37 &\small 22 &\small 21&\small 15.83 &\small 15.11 &\small 1 &\small  0.72 &\centering \small  901 \tabularnewline 
	 \small TOUTV  &\small 169 &\small 86  &\small 50.89 &\small   8.37 &\small  25.04 &\small  13.75 &\small 20 &\small 21&\small 11.83 &\small 12.43 &\small -1&\small -0.59 &\centering \small  873 \tabularnewline 
	 \small NTRAN  &\small 151 &\small 81  &\small 53.64 &\small   7.35 &\small  25.83 &\small  13.91 &\small 23 &\small 21&\small 15.23 &\small 13.91 &\small 2 &\small  1.32 &\centering \small  695 \tabularnewline 
	 \small AVBLS  &\small 183 &\small 93  &\small 50.82 &\small   5.66 &\small  23.12 &\small  12.38 &\small 25 &\small 22&\small 13.66 &\small 12.02 &\small 3 &\small  1.64 &\centering \small  677 \tabularnewline 
	 \small NTRBL  &\small 149 &\small 80  &\small 53.69 &\small   1.61 &\small  22.76 &\small  17.60 &\small 28 &\small 21&\small 18.79 &\small 14.09 &\small 7 &\small  4.70 &\centering \small  222 \tabularnewline 
	 \small ATRCT  &\small 188 &\small 86  &\small 45.74 &\small  -2.4  &\small  28.21 &\small  37.76 &\small 15 &\small 23&\small  7.98 &\small 12.23 &\small -8&\small -4.26 &\centering \small -790 \tabularnewline 
      \hline
	
	\end{tabular}
	}
	\caption{Aggregated results using blockchain ribbons for long and short operations.}

    \label{table2}

\end{table}

Table \ref{table3} shows the results in long operations. The total percentage of winning trades is $58.13\%$, in contrast to $50.49\%$ when the type of operation is not differentiated. Moreover, all the blockchain metrics have positive values in both Trade Profit and Strategy Balance. This indicates that there is a greater statistical advantage when long and short trades positions are disaggregated.

\begin{table}
 	\tiny
  	\renewcommand{\arraystretch}{1.3} 
  
  	\resizebox{1\textwidth}{!} {

	\begin{tabular}{|>{\centering}p{1.3cm}|c|>{\centering}p{1.0cm}|>{\centering}p{1.1cm}|>{\centering}p{1.0cm}|>{\centering}p{1.25cm}|>{\centering}p{1.15cm}|>{\centering}p{1.15cm}|>{\centering}p{1.17cm}|>{\centering}p{1.12cm}|>{\centering}p{1.18cm}|>{\centering}p{1.0cm}|>{\centering}p{1.09cm}|p{1.16cm}|}

      	\hline
	
	\textbf{Metric}			& 
      	\textbf{Trades}			& 
      	\textbf{Winning Trades} 		& 
      	\textbf{Winning Trades (\%) }	& 
      	\textbf{Trade Profit (\%)} 			& 
      	\textbf{Maximum Achieved (\%)}		& 
      	\textbf{Minimum BT Price (\%)}			& 
      	\textbf{Threshold achieved} 			&	
     	\textbf{Stop Loss activated}			&
      	\textbf{Threshold achieved (\%)}		& 
      	\textbf{Stop Loss activated (\%)}		&	
      	\textbf{Strategy Balance}				&	
	\textbf{Trades Total Balance Strategy (\%)}	&
	\centering  \textbf{Strategy Total Profit (\%)}  \arraybackslash \\
	\hline
	\small   &\small  &\small   &\small  &\small  &\small  &\small  &\small  &\small  &\small  &\small   &\small &\small &\centering \small  \tabularnewline
	 \small NADDU  &\small 64  &\small 37 &\small 57.81 &\small 117.41 &\small 274.13 &\small   7.50 &\small 19 &\small 6 &\small 29.69 &\small  9.38 &\small 13 &\small 20.31 &\centering \small 4905 \tabularnewline 
	 \small CPTRA  &\small 92  &\small 60 &\small 65.22 &\small  64.26 &\small 133.71 &\small   6.76 &\small 34 &\small 2 &\small 36.96 &\small  2.17 &\small 32 &\small 34.78 &\centering \small 4467 \tabularnewline 
	 \small TRFUS  &\small 59  &\small 35 &\small 59.32 &\small  86.54 &\small 201.73 &\small   7.21 &\small 14 &\small 3 &\small 23.73 &\small  5.08 &\small 11 &\small 18.64 &\centering \small 3906 \tabularnewline 
	 \small TRFEE  &\small 61  &\small 34 &\small 55.74 &\small  85.70 &\small 176.10 &\small  10.75 &\small 13 &\small 5 &\small 21.31 &\small  8.20 &\small  8 &\small 13.11 &\centering \small 3864 \tabularnewline 
	 \small TRVOU  &\small 70  &\small 42 &\small 60.00 &\small  47.32 &\small 100.37 &\small   6.56 &\small 14 &\small 4 &\small 20.00 &\small  5.71 &\small 10 &\small 14.29 &\centering \small 2691 \tabularnewline 
	 \small ETRVU  &\small 78  &\small 38 &\small 48.72 &\small  39.99 &\small  88.58 &\small   7.68 &\small 16 &\small 5 &\small 20.51 &\small  6.41 &\small 11 &\small 14.10 &\centering \small 2532 \tabularnewline 
	 \small MKPRU  &\small 39  &\small 24 &\small 61.54 &\small  93.14 &\small 188.16 &\small   6.88 &\small 16 &\small 1 &\small 41.03 &\small  2.56 &\small 15 &\small 38.46 &\centering \small 2460 \tabularnewline 
	 \small MIREV  &\small 47  &\small 28 &\small 59.57 &\small  66.90 &\small 138.03 &\small   6.14 &\small 17 &\small 1 &\small 36.17 &\small  2.13 &\small 16 &\small 34.04 &\centering \small 2373 \tabularnewline 
	 \small MKTCP  &\small 36  &\small 20 &\small 55.56 &\small 105.81 &\small 214.44 &\small   7.69 &\small 17 &\small 1 &\small 47.22 &\small  2.78 &\small 16 &\small 44.44 &\centering \small 2349 \tabularnewline 
	 \small ETRAV  &\small 114 &\small 69 &\small 60.53 &\small  20.33 &\small  33.00 &\small   6.51 &\small 21 &\small 5 &\small 18.42 &\small  4.39 &\small 16 &\small 14.04 &\centering \small 2155 \tabularnewline 
	 \small TOUTV  &\small 85  &\small 46 &\small 54.12 &\small  24.21 &\small  40.47 &\small   7.47 &\small 16 &\small 5 &\small 18.82 &\small  5.88 &\small 11 &\small 12.94 &\centering \small 1795 \tabularnewline 
	 \small NTREP  &\small 70  &\small 40 &\small 57.14 &\small  32.16 &\small  70.95 &\small   8.09 &\small 18 &\small 8 &\small 25.71 &\small 11.43 &\small 10 &\small 14.29 &\centering \small 1645 \tabularnewline 
	 \small NTRAN  &\small 76  &\small 48 &\small 63.16 &\small  25.49 &\small  44.11 &\small   7.20 &\small 19 &\small 6 &\small 25.00 &\small  7.89 &\small 13 &\small 17.11 &\centering \small 1614 \tabularnewline 
	 \small AVBLS  &\small 92  &\small 55 &\small 59.78 &\small  18.55 &\small  39.55 &\small   7.63 &\small 23 &\small 6 &\small 25.00 &\small  6.52 &\small 17 &\small 18.48 &\centering \small 1587 \tabularnewline 
	 \small NTRBL  &\small 75  &\small 46 &\small 61.33 &\small  19.02 &\small  36.91 &\small   8.16 &\small 22 &\small 6 &\small 29.33 &\small  8.00 &\small 16 &\small 21.33 &\centering \small 1299 \tabularnewline 
	 \small HRATE  &\small 21  &\small 12 &\small 57.14 &\small 176.04 &\small 612.66 &\small 176.18 &\small 12 &\small 4 &\small 57.14 &\small 19.05 &\small  8 &\small 38.10 &\centering \small 1099 \tabularnewline 
	 \small ATRCT  &\small 94  &\small 45 &\small 47.87 &\small  10.80 &\small  51.30 &\small  43.65 &\small 13 &\small 7 &\small 13.83 &\small  7.45 &\small  6 &\small  6.38 &\centering \small  885 \tabularnewline 
	 \small DIFF   &\small 14  &\small 11 &\small 78.57 &\small 211.76 &\small 837.86 &\small 263.01 &\small 10 &\small 4 &\small 71.43 &\small 28.57 &\small  6 &\small 42.86 &\centering \small  166 \tabularnewline

      \hline
	
	\end{tabular}
	}
		
	\caption{Results for long operations.}

    \label{table3}

\end{table}

We also see in Table \ref{table3} that seven indicators exceed $60\%$ in Winning Trades; in particular, DIFF has a return of $78\%$ of winning trades, although the number of signals is low. However, applying the $30\%$ strategy at 1:1, the NADDU and CPTRA metrics show the best performance as measured by the indicators Trade Profit, Strategy Balance and Total Profit of the Strategy. This is because they generate a much higher number of trades, although their success rate is lower. This time there are six metrics that outperform the Bitcoin price regarding the Total Profit. 

\begin{table}
 	\tiny
  	\renewcommand{\arraystretch}{1.3} 
  
  	\resizebox{1\textwidth}{!} {

	\begin{tabular}{|>{\centering}p{1.3cm}|c|>{\centering}p{1.0cm}|>{\centering}p{1.1cm}|>{\centering}p{1.0cm}|>{\centering}p{1.25cm}|>{\centering}p{1.15cm}|>{\centering}p{1.15cm}|>{\centering}p{1.17cm}|>{\centering}p{1.12cm}|>{\centering}p{1.18cm}|>{\centering}p{1.0cm}|>{\centering}p{1.09cm}|p{1.16cm}|}

      	\hline
	
	\textbf{Metric}			& 
      	\textbf{Trades}			& 
      	\textbf{Winning Trades} 		& 
      	\textbf{Winning Trades (\%) }	& 
      	\textbf{Trade Profit (\%)} 			& 
      	\textbf{Maximum Achieved (\%)}		& 
      	\textbf{Minimum BT Price (\%)}			& 
      	\textbf{Threshold achieved} 			&	
     	\textbf{Stop Loss activated}			&
      	\textbf{Threshold achieved (\%)}		& 
      	\textbf{Stop Loss activated (\%)}		&	
      	\textbf{Strategy Balance}				&	
	\textbf{Trades Total Balance Strategy (\%)}	&
	\centering  \textbf{Strategy Total Profit (\%)}  \arraybackslash \\
	\hline
	\small   &\small  &\small   &\small  &\small  &\small  &\small  &\small  &\small  &\small  &\small   &\small &\small &\centering \small  \tabularnewline
	 \small MKTCP  &\small 35  &\small 12 &\small 34.29 &\small -3.30 &\small 16.53 &\small 13.25 &\small 10 &\small 4  &\small 28.57 &\small 11.43 &\small 6   &\small 17.14 &\centering \small    76 \tabularnewline 
	 \small NADDU  &\small 63  &\small 28 &\small 44.44 &\small -0.79 &\small  9.63 &\small  7.48 &\small 6  &\small 3  &\small  9.52 &\small  4.76 &\small 3   &\small  4.76 &\centering \small   -16 \tabularnewline 
	 \small MKPRU  &\small 38  &\small 12 &\small 31.58 &\small -3.88 &\small 15.27 &\small 12.84 &\small 8  &\small 4  &\small 21.05 &\small 10.53 &\small 4   &\small 10.53 &\centering \small   -19 \tabularnewline 
	 \small HRATE  &\small 20  &\small 10 &\small 50.00 &\small -2.50 &\small 12.08 &\small 12.86 &\small 3  &\small 3  &\small 15.00 &\small 15.00 &\small 0   &\small  0.00 &\centering \small   -55 \tabularnewline 
	 \small DIFF   &\small 13  &\small  6 &\small 46.15 &\small 11.63 &\small 14.36 &\small 18.70 &\small 2  &\small 3  &\small 15.38 &\small 23.08 &\small -1  &\small -7.69 &\centering \small  -136 \tabularnewline 
	 \small MIREV  &\small 46  &\small 18 &\small 39.13 &\small -5.06 &\small 13.75 &\small 13.59 &\small 9  &\small 7  &\small 19.57 &\small 15.22 &\small 2   &\small  4.35 &\centering \small  -138 \tabularnewline 
	 \small TRVOU  &\small 69  &\small 29 &\small 42.03 &\small -2.10 &\small 11.77 &\small 13.11 &\small 9  &\small 8  &\small 13.04 &\small 11.59 &\small 1   &\small  1.45 &\centering \small  -148 \tabularnewline 
	 \small ETRVU  &\small 77  &\small 35 &\small 45.45 &\small -2.75 &\small  9.07 &\small 11.05 &\small 8  &\small 7  &\small 10.39 &\small  9.09 &\small 1   &\small  1.30 &\centering \small  -218 \tabularnewline 
	 \small TRFUS  &\small 58  &\small 26 &\small 44.83 &\small -7.75 &\small 11.35 &\small 14.05 &\small 8  &\small 10 &\small 13.79 &\small 17.24 &\small -2  &\small -3.45 &\centering \small  -428 \tabularnewline 
	 \small CPTRA  &\small 90  &\small 34 &\small 37.78 &\small -8.46 &\small 13.73 &\small 19.17 &\small 14 &\small 14 &\small 15.56 &\small 15.56 &\small 0   &\small  0.00 &\centering \small  -614 \tabularnewline 
	 \small TRFEE  &\small 61  &\small 27 &\small 44.26 &\small 13.49 &\small 11.74 &\small 24.43 &\small 8  &\small 14 &\small 13.11 &\small 22.95 &\small -6  &\small -9.84 &\centering \small  -767 \tabularnewline 
	 \small ETRAV  &\small 113 &\small 53 &\small 46.90 &\small -5.50 &\small  6.30 &\small 15.57 &\small 4  &\small 11 &\small  3.54 &\small  9.73 &\small -7  &\small -6.19 &\centering \small  -862 \tabularnewline 
	 \small TOUTV  &\small 84  &\small 40 &\small 47.62 &\small -7.65 &\small  9.42 &\small 20.11 &\small 4  &\small 16 &\small  4.76 &\small 19.05 &\small -12 &\small-14.29 &\centering \small  -934 \tabularnewline 
	 \small NTREP  &\small 69  &\small 24 &\small 34.78 &\small 11.44 &\small  6.75 &\small 18.71 &\small 4  &\small 13 &\small  5.80 &\small 18.84 &\small -9  &\small-13.04 &\centering \small  -934 \tabularnewline 
	 \small NTRAN  &\small 75  &\small 33 &\small 44.00 &\small 11.03 &\small  7.30 &\small 20.71 &\small 4  &\small 15 &\small  5.33 &\small 20.00 &\small -11 &\small-14.67 &\centering \small -1023 \tabularnewline 
	 \small AVBLS  &\small 91  &\small 38 &\small 41.76 &\small -7.37 &\small  6.52 &\small 17.18 &\small 2  &\small 16 &\small  2.20 &\small 17.58 &\small -14 &\small-15.38 &\centering \small -1049 \tabularnewline 
	 \small NTRBL  &\small 74  &\small 34 &\small 45.95 &\small 16.04 &\small  8.42 &\small 27.17 &\small 6  &\small 15 &\small  8.11 &\small 20.27 &\small -9  &\small-12.16 &\centering \small -1194 \tabularnewline 
	 \small ATRCT  &\small 94  &\small 41 &\small 43.62 &\small 15.63 &\small  5.12 &\small 31.87 &\small 2  &\small 16 &\small  2.13 &\small 17.02 &\small -14 &\small-14.89 &\centering \small -1702 \tabularnewline

       \hline
	
	\end{tabular}
	}
		
	\caption{Results for short operations.}

    \label{table4}

\end{table}

Grouping by short operations, the total percentage of winning trades is 42.75\% (Table \ref{table4}). This indicates that in short trades there is no statistical advantage in the number of winning trades versus the number of open trades over time. The only blockchain metric that has 50\% winning trades is HRATE.

Moreover, in short operations, all the metrics present negative values in "Trade Profit" and in "Total Profit Strategy" measures. The only indicator that presents a positive Total Profit is MKTCP, but with a value of 76\%.


\subsection{Further considerations }

\label{s:Results.4.2}

The previous analysis has been carried out from the birth of Bitcoin in 2009, when its value was zero, until July 2022, when it value was close to \$20,000. Therefore, one might conclude that long signals are more profitable than short signals because the Bitcoin has appreciated over time.

To rule out this hypothesis, the data was filtered from December 2017, when the value of the Bitcoin was \$20,000, until July 2022, when its value was the same. Long trading results in that period remain positive, although with much smaller profits, while the ranking changes slightly: CPTRA moves to the first position but the top performers stand. In the case of short trading, however, all metrics, except again CPTRA, have negative performances.

Therefore, the hypothesis that long signals are more profitable than short signals due to the appreciation of the Bitcoin since its inception is refused. It is also worth highlighting that the CPTRA metric records the best performance in both long and short operations in the period December 2017 - July 2022.

From the above analysis with historical data, we may conclude that the use of blockchain metrics increases the profitability in long operations. As a drawback or opportunity for improvement, we also conclude that they (including the Bitcoin price) are not effective in short operations. Therefore, when it comes to closing trades and making profits or opening short positions, it is necessary to look for other options. 


\section{Improving on the CPTRA and monotonic metrics}

\label{s:applications}

In this section we focus on four particular blockchain metrics: CPTRA, MWNUS, NTRAT and BLCHS. 

First, we apply linear regression to the monotonic maxima of the CPTRA (Bitcoin Cost per Transaction) metric, see Figure \ref{figure7}, to define an \textquotedblleft adjusted\textquotedblright\ metric that has also a good performance in short operations too.

Second, we adapt the MWNUS (My Wallet Number of Users), NTRAT (Total Number of Transactions) and BLCHS (Api Block-Chain Size) metrics to the ribbon technique. To overcome the monotonicity of these three metrics, we use time derivatives.


\subsection{\emph{An improved CPTRA metric}}

\label{s:applications.1}

We hypothesized in Section \ref{s:blockchain ribbons.3.4} that the line drawn through the ever higher maxima of the CPTRA ribbon, Figure \ref{figure7}, can be interpreted as resistance and support zones in which there is a strong rejection towards the opposite price direction of the Bitcoin. We build on this here.


\subsubsection{\emph{ Linear Regression of Maximum and Minimum CPTRA}}

\label{s:applications.1.1}

To construct the regression line through historical highs that exceed the previous ones, a time series is created with the maximum value of the CPTRA moving averages of a period of 30 days in the years 2012 to 2022, whenever a yearly maximum exceeds the previous one. That is, if a local maximum does not exceed a previous local maximum, it is not entered into the regression equation. 

Therefore, set

\begin{equation}
y_{k} = \max \{y(t)\mbox{ in the $k$th year}\} 
\end{equation}%

\noindent so that $y_{k}$ is the maximum of the CPTRA SMA-30 daily values $y(t)$ in the year $2011+k$, $1\leq k\leq 11$ (see Figure \ref{figure7}). Now, define recursively $Y_{i}=y_{j(i)}$, $i\geq 1$, where $j(1)=1$ and 

\begin{equation}
j(i)=\min \{k>j(i-1):y_{k}>Y_{i-1}\}
\end{equation}

\noindent for $i\geq 2$. In view of Figure \ref{figure7}, we have: $Y_{1}=y_{1}$ (year 2012), $Y_{2}=y_{3}$ (year 2014), $Y_{3}=y_{7}$ (year 2018), $Y_{4}=y_{10}$ (year 2021), and $Y_{5}=y_{11}$ (year 2022)  and $Y_{6}=y_{12}$ (year 2022). 

\begin{figure}[th]
\centering
\includegraphics[width=110mm]{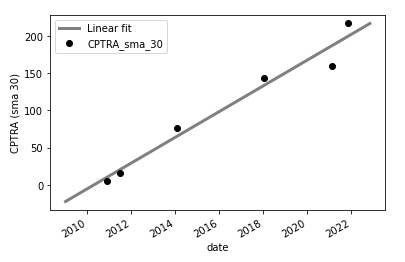}
\caption{ Linear regression of the CPTRA monotonic maxima. }
\label{figure10}
\end{figure}

The line drawn in Figure \ref{figure7} is the regression line through the points $(i,Y_{i})$, $1\leq i\leq 5$. We call the values $Y_{i}$ the CPTRA monotonic maxima. As shown in Figure \ref{figure10}, the CPTRA monotonic maxima turn out to be arranged almost perfectly along the regression line.

\begin{figure}[th]
\centering
\includegraphics[width=110mm]{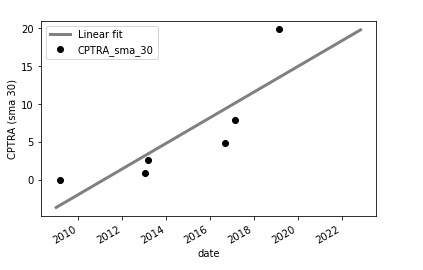}
\caption{ Linear regression of the CPTRA minima. }
\label{figure11}
\end{figure}

To obtain a similar regression line for the minimum values, a series was created with the historical minima of the CPTRA SMA-30 daily values. This time, the condition that a minimum had to exceed the previous minimum was not taken into account because the curve of the minima is more oscillating than that of the maxima. The resulting regression line is shown in Figure \ref{figure11}.

\subsubsection{\emph{The Adjusted CPTRA metric}}

\label{s:applications.1.2}

\begin{figure}[th]
\centering
\includegraphics[width=130mm]{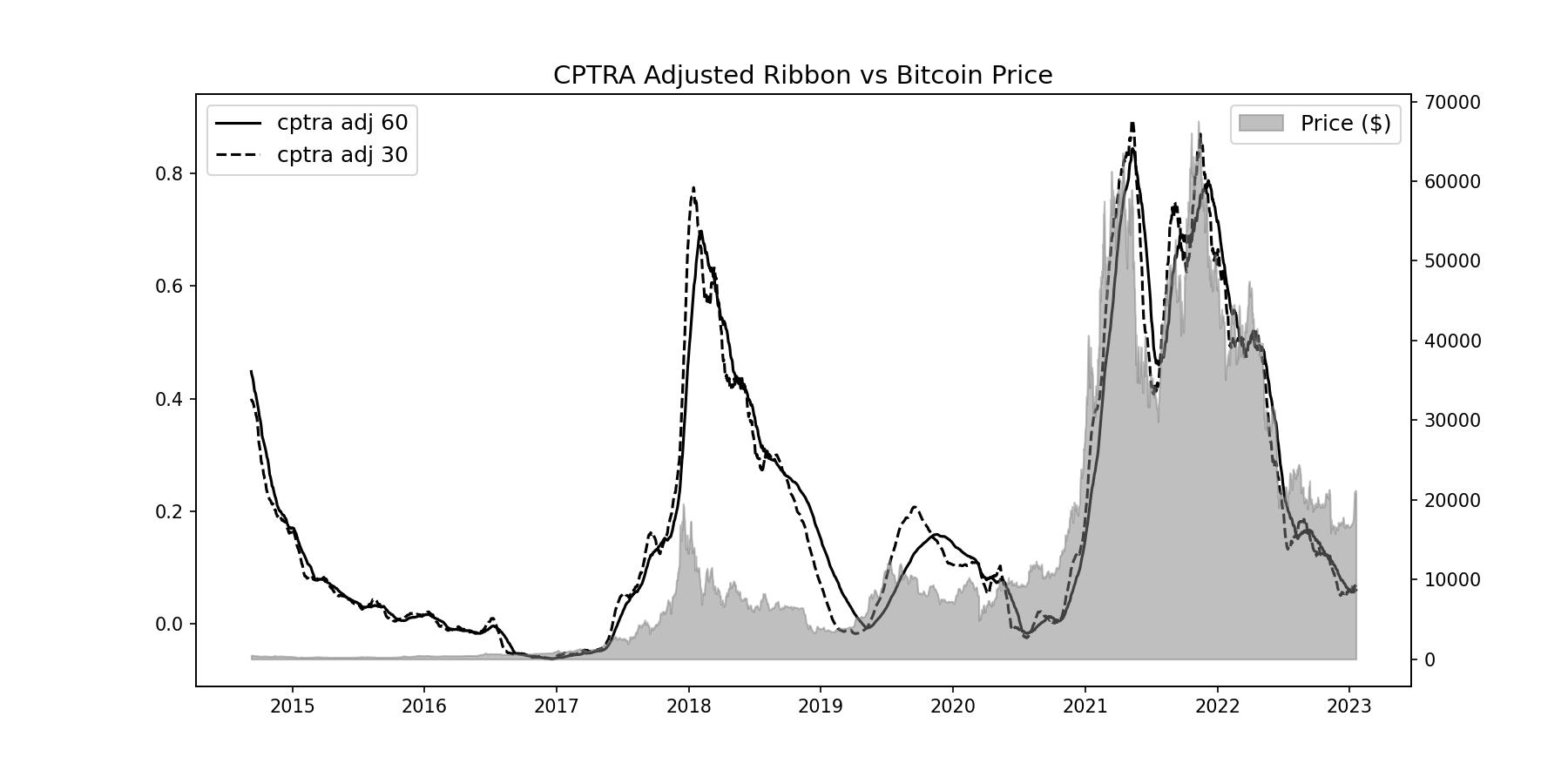}
\caption{ The Adjusted CPTRA ribbon.}
\label{figure12}
\end{figure}

The new Adjusted CPTRA (AdCPTRA) metric is defined as

\begin{equation}
AdCPTRA(i)=\frac{SMA(i)-MinLR(i)}{MaxLR(i)}
\end{equation}%
where $SMA(i)$ is the moving average of CPTRA of period 30 at day $i$ and $MaxLR(i)$ (resp. $MinLR(i)$) is the value of the linear regression of the CPTRA monotonic maxima (resp. minima) obtained as described above. The result is a graph, see Figure \ref{figure12}, with two moving averages of 30 and 60 days that oscillate between 0 and 1.

By implementing the new ribbon, two areas can be distinguished in Figure \ref{figure12}. In the upper zone, where AdCPTRA exceeds 0.6, we see a strong rejection in the price, as well as a decline in both the price and the metric. In the lower zone, where the value of the indicator lies below 0.2 and there are long signals, we see strong increases in the price.

\subsubsection{Performance of the Adjusted CPTRA ribbon}

\label{s:applications.1.3}

To check the performance of the AdCPTRA ribbon, the same simulations for long and short signals as in Section \ref{s:methods.4} were carried out, this time with two additional provisos: For long signals, the AdCPTRA metric must be less than 0.3, while for short signals it must be greater than 0.6. The results are shown in Table \ref{table5}.

\begin{table}
 	\tiny
  	\renewcommand{\arraystretch}{1.3} 
  
  	\resizebox{1\textwidth}{!} {

	\begin{tabular}{|>{\centering}p{1.6cm}|>{\centering}p{0.7cm}|c|>{\centering}p{1.0cm}|>{\centering}p{1.1cm}|>{\centering}p{1.0cm}|>{\centering}p{1.25cm}|>{\centering}p{1.15cm}|>{\centering}p{1.15cm}|>{\centering}p{1.17cm}|>{\centering}p{1.12cm}|>{\centering}p{1.18cm}|>{\centering}p{1.0cm}|>{\centering}p{1.09cm}|p{1.16cm}|}

      	\hline
	
	\textbf{Metric}			& 
      	\textbf{Opt.}			&
      	\textbf{Trades}			& 
      	\textbf{Winning Trades} 		& 
      	\textbf{Winning Trades (\%) }	& 
      	\textbf{Trade Profit (\%)} 			& 
      	\textbf{Maximum Achieved (\%)}		& 
      	\textbf{Minimum BT Price (\%)}			& 
      	\textbf{Threshold achieved} 			&	
     	\textbf{Stop Loss activated}			&
      	\textbf{Threshold achieved (\%)}		& 
      	\textbf{Stop Loss activated (\%)}		&	
      	\textbf{Strategy Balance}				&	
	\textbf{Trades Total Balance Strategy (\%)}	&
	\centering  \textbf{Strategy Total Profit (\%)}  \arraybackslash \\
	\hline
	 \small  &\small   &\small   &\small  &\small  &\small  &\small &\small  &\small  &\small    &\small  &\small  &\small  &\small  &\centering \small \tabularnewline 
	 \small AdCPTRA &\small long  &\small 32  &\small 23 &\small 71.88 &\small 92.11 &\small 184.60 &\small  9.43 &\small 15 &\small  2  &\small 46.88 &\small  6.25 &\small 13 &\small 40.63 &\centering \small 1740 \tabularnewline 
	 \small CPTRA   &\small long  &\small 92  &\small 60 &\small 65.22 &\small 64.26 &\small 133.71 &\small  6.76 &\small 34 &\small  2  &\small 36.96 &\small  2.17 &\small 32 &\small 34.78 &\centering \small 4467 \tabularnewline 
	 \small AdCPTRA &\small short &\small 6   &\small  4 &\small 66.67 &\small 12.01 &\small  38.11 &\small 12.31 &\small 4  &\small  1  &\small 66.67 &\small 16.67 &\small  3 &\small 50.00 &\centering \small   96 \tabularnewline 
	 \small CPTRA   &\small short &\small 90  &\small 34 &\small 37.78 &\small -8.46 &\small  13.73 &\small 19.17 &\small 14 &\small 14  &\small 15.56 &\small 15.56 &\small  0 &\small  0.00 &\centering \small -614 \tabularnewline 

       \hline
	
	\end{tabular}
	}
		
	\caption{Performance of the CPTRA ribbon vs the Adjusted CPTRA ribbon. The CPTRA data is reproduced from Table \ref{table3} (long operation) and Table \ref{table4} (short operation) to facilitate the comparison with the AdCPTRA data.}

    \label{table5}

\end{table}

We see in Table \ref{table5} that, in both long and short signals, the AdCPTRA ribbon outperforms the CPTRA ribbon on most indicators. This is a consequence of the reduction in the number of signals. 

Specifically, in long signals the number of trades drops from 92 to 32, and the percentages of Winning Trades, Trade Profit and the Average of the Maximum Bitcoin Price improve. However, the Strategy Total Profit drops from $ 4467\%$ to $ 1740\%$ (in total $ 2727\%$). Therefore, the use of the AdCPTRA metric to long signals improves the performance but reduces the profit.

Regarding the short signals, the number of trades drops drastically from 90 to 6, but the percentage of Winning Trades increases from $37\%$ to $66\%$. More importantly, the Strategy Total Profit rises from $-614\%$ to $96\%$, which confirms that AdCPTRA provides a statistical advantage on short signals.

\subsection{\emph{An improved MWNUS metric}}

\label{s:applications.2}

Although MWNUS (Bitcoin My Wallet Number of Users) has a good correlation with the Bitcoin price, its value increases monotonically, as shown in Figure \ref{figure4}. This means that its ribbon lines do not cross. This being the case, we will use time derivatives together with moving averages of 10 and 20 days, the reason for these shorter periods being that increases in the number of wallets is a consequence of a sudden change in trend. As a result, the price oscillations have a shorter time frame.

Specifically, the Adjusted MWNUS (AdMWNUS) indicator is defined as 
\begin{equation}
AdMWNUS(t)=\frac{1}{T}\sum_{i=0}^{T-1}x^{\prime }(t-i)  \label{AMWNUS}
\end{equation}%
where $x(t)=MWNUS(t)$, $x^{\prime }(t)$ is the time derivative of $x(t)$ and $T=10$ (for SMA-10) or $20$ (for SMA-20). The time derivatives are calculated by backward differences.

\begin{figure}[th]
\centering
\includegraphics[width=130mm]{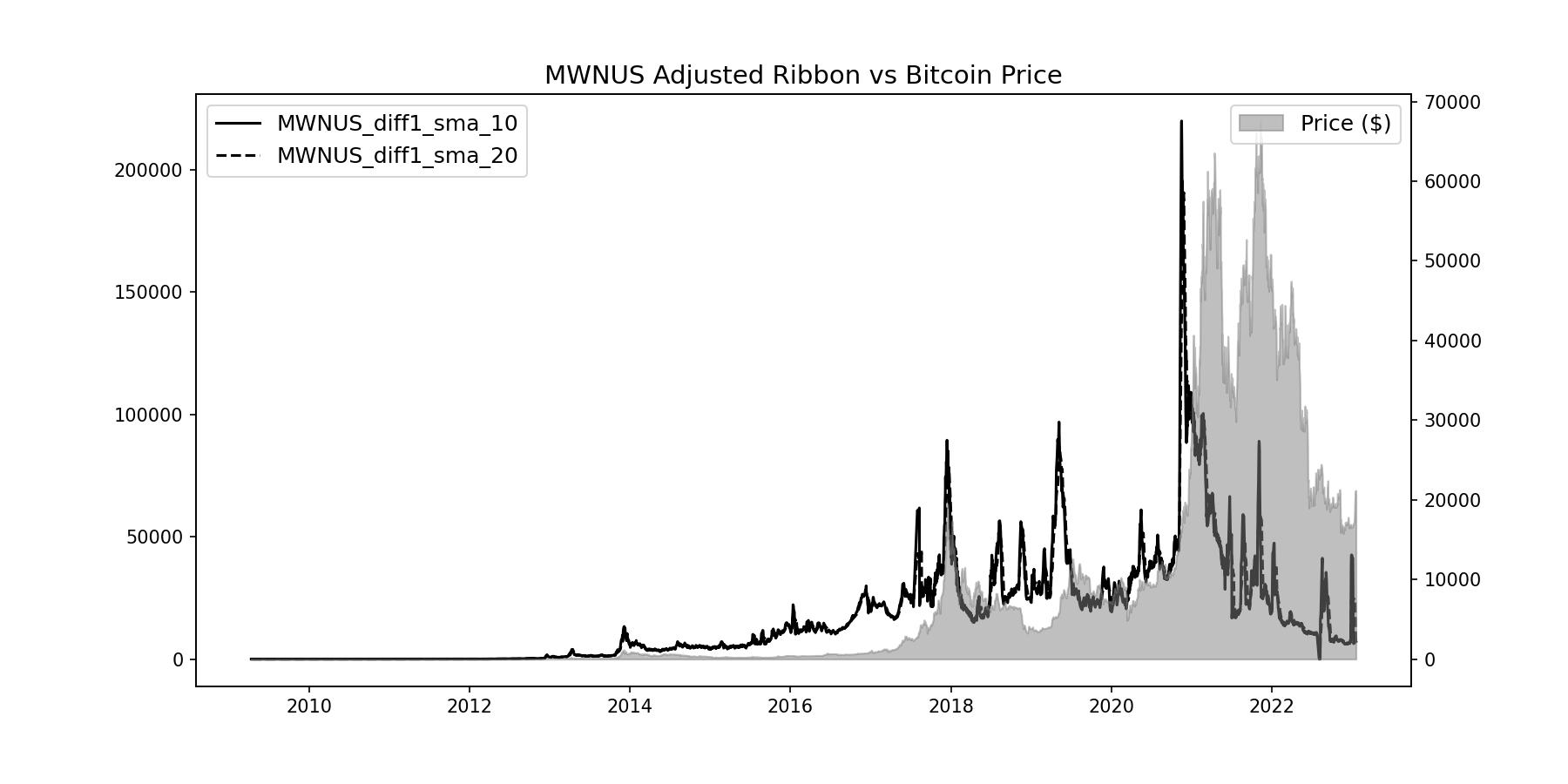}
\caption{ The Adjusted MWNUS ribbon.}
\label{figure13}
\end{figure}

Figure \ref{figure13} shows that the ribbon lines of the AdMWNUS metric reproduce the oscillations of the Bitcoin price.

\subsubsection{\emph{Performance of the Adjusted MWNUS ribbon}}

\label{s:applications.4.3}

To analyze the performance of the AdMWNUS ribbon, the same simulations were made with long and short signals as in Section \ref {s:methods.4}. However, since in this case the swings are larger (due to the shorter periods of 10 and 20 days), the target (profit and risk margin) has been adjusted to 10\%. The results are shown in Table \ref{table6}, along with other indicators to be introduced in the following sections.

\begin{table}
 	\tiny
  	\renewcommand{\arraystretch}{1.3} 
  
  	\resizebox{1\textwidth}{!} {

	\begin{tabular}{|>{\centering}p{1.7cm}|>{\centering}p{0.6cm}|c|>{\centering}p{1.0cm}|>{\centering}p{1.1cm}|>{\centering}p{1.0cm}|>{\centering}p{1.25cm}|>{\centering}p{1.15cm}|>{\centering}p{1.15cm}|>{\centering}p{1.17cm}|>{\centering}p{1.12cm}|>{\centering}p{1.18cm}|>{\centering}p{1.0cm}|>{\centering}p{1.09cm}|p{1.16cm}|}

      	\hline
	
	\textbf{Metric}			& 
      	\textbf{Opt.}			&
      	\textbf{Trades}			& 
      	\textbf{Winning Trades} 		& 
      	\textbf{Winning Trades (\%) }	& 
      	\textbf{Trade Profit (\%)} 			& 
      	\textbf{Maximum Achieved (\%)}		& 
      	\textbf{Minimum BT Price (\%)}			& 
      	\textbf{Threshold achieved} 			&	
     	\textbf{Stop Loss activated}			&
      	\textbf{Threshold achieved (\%)}		& 
      	\textbf{Stop Loss activated (\%)}		&	
      	\textbf{Strategy Balance}				&	
	\textbf{Trades Total Balance Strategy (\%)}	&
	\centering  \textbf{Strategy Total Profit (\%)}  \arraybackslash \\
	\hline
	 \small  &\small   &\small   &\small  &\small  &\small   &\small  &\small  &\small  &\small   &\small  &\small  &\small  &\small  &\centering \small  \tabularnewline 
	 \small AdBLCHS &\small long  &\small 264  &\small 148 &\small 56.06 &\small  1.06 &\small   3.80 &\small  2.65 &\small 22 &\small 17  &\small  8.33 &\small  6.44 &\small  5 &\small  1.89 &\centering \small  124 \tabularnewline 
	 \small AdBLCHS &\small short &\small 264  &\small 137 &\small 51.89 &\small  0.13 &\small   3.08 &\small  2.80 &\small 20 &\small 20  &\small  7.58 &\small  7.58 &\small  0 &\small  0.00 &\centering \small -234 \tabularnewline 
	 \small AdNTRAT &\small long  &\small 116  &\small 65  &\small 56.03 &\small  2.42 &\small   6.93 &\small  3.61 &\small 24 &\small 12  &\small 20.69 &\small 10.34 &\small 12 &\small 10.34 &\centering \small  438 \tabularnewline 
	 \small AdNTRAT &\small short &\small 112  &\small 63  &\small 56.25 &\small  0.20 &\small   5.14 &\small  4.76 &\small 20 &\small 12  &\small 17.86 &\small 10.71 &\small  8 &\small  7.14 &\centering \small  144 \tabularnewline 
   \small AdMWNUS &\small long  &\small 117  &\small 66  &\small 56.41 &\small  2.46 &\small   6.65 &\small  3.58 &\small 25 &\small 12  &\small 21.37 &\small 10.26 &\small 13 &\small 11.11 &\centering \small  469 \tabularnewline
   \small AdMWNUS &\small short &\small 118  &\small 64  &\small 54.24 &\small  0.07 &\small   5.23 &\small  5.03 &\small 21 &\small 14  &\small 17.80 &\small 11.86 &\small  7 &\small  5.93 &\centering \small   98 \tabularnewline
   \small AdCPTRA &\small long  &\small 32   &\small 23  &\small 71.88 &\small 92.11 &\small 184.60 &\small  9.43 &\small 15 &\small  2  &\small 46.88 &\small  6.25 &\small 13 &\small 40.63 &\centering \small 1740 \tabularnewline
   \small AdCPTRA &\small short &\small 6    &\small 4   &\small 66.67 &\small 12.01 &\small  38.11 &\small 12.31 &\small  4 &\small  1  &\small 66.67 &\small 16.67 &\small  3 &\small 50.00 &\centering \small   96 \tabularnewline

       \hline
	
	\end{tabular}
	}
		
	\caption{Performance of the adjusted ribbons.}

    \label{table6}

\end{table}

In this case, the Percentage of Winning Trades on short signals has risen to $54\%$, but the expected statistical advantage has not been reached. In addition, the Total Profit on short signals is now $56\%$, while on long signals the Total Profit is $469\%$. So we conclude that the AdMWNUS metric does bring a statistical advantage on long signals also in periods when prices are stable, as do the metrics analyzed in previous sections.


\subsection{\emph{An improved NTRAT metric}}

\label{s:applications.3}

As can be seen in Figure \ref{figure5}, the NTRAT indicator contains accumulated values because it refers to the size of the devices that have connected to the Blockchain API.

As before, to define an Adjusted NTRAT (AdNTRAT) metric, we use moving averages of 10 and 20 days to smooth out the series, time derivatives to obtain fluctuations, and crossovers of the ribbon lines to locate the buy and sell signals. Thus, the Adjusted NTRAT (AdNTRAT) metric is defined as 

\begin{equation}
ANTRAT(t)=\frac{1}{T}\sum_{i=0}^{T-1}x^{\prime }(t-i)  \label{ANTRAT}
\end{equation}%
where $x(t)=NTRAT(t)$, $x^{\prime }(t)$ is the time derivative of $x(t)$, and $T=10$ (SMA-10) or $20$ (SMA-20). The time derivatives are calculated by backward differences.

\begin{figure}[th]
\centering
\includegraphics[width=130mm]{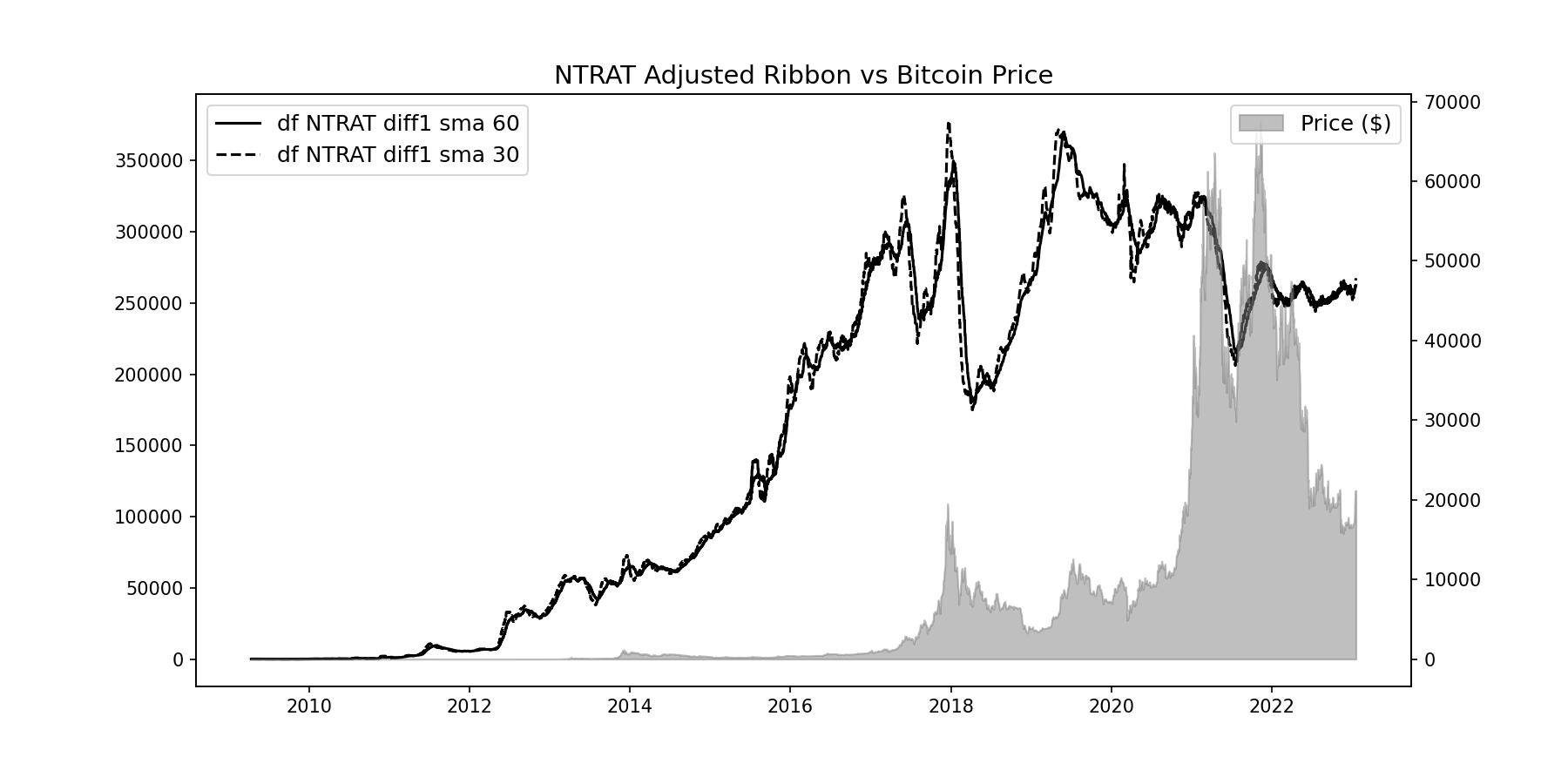}
\caption{ The Adjusted NTRAT ribbon.}
\label{figure14}
\end{figure}

Figure \ref{figure14} shows that the long and short signals obtained from the Adjusted NTRAT metric reproduce the oscillations of the Bitcoin price.

\subsubsection{\emph{Performance of the Adjusted NTRAT ribbon}}

\label{s:applications.4ntrat}

To analyze the performance of the Adjusted NTRAT ribbon, the same simulations were made as in Section \ref {s:applications.2}. The results are shown in Table \ref{table6}.

In sum, the performances of the AdNTRAT ribbon for long and short operations get closer, namely: 56\% and 56\% in the respective Winning Trades, and 438\% and 144\% in the respective Total Profits. Therefore, this indicator provides more statistical advantage and could help especially in short signals.


\subsection{\emph{An improved BLCHS metric}}

\label{s:applications.4}

Similarly to AdMWNUS and AdNTRAT, the Adjusted BLCHS metric (AdBLCHS) is defined via derivatives and the corresponding ribbon via moving averages of 10 and 20 days. 

According to Table \ref{table6}, the Percentage of Winning Trades on long signals is $56\%$, but the Percentage of Winning Trades on short signals is $52\%$, so the expected statistical advantage is not reached. In addition, the Total Profit on short signals is now $-234\%$, while the Total Profit on long signals is $124\%$. So we conclude that the AdBLCHS metric does not bring a statistical advantage on long signals (including periods when prices are stable), contrarily to AdMWNUS, AdNTRAT, and also AdCPTRA.


\section{Performance of the blockchain metrics using predictive models}

\label{s:predictions}

Once we have measured the performance of the blockchain metrics in detecting long and short positions in Sections 4 and 5, now we test how the blockchain metrics perform when it comes to make predictions. To this end we are going first to determine what kind of predictive model and what handling of the variables could be more suitable to build an algorithmic trading system based on Machine Learning. Thereby, only data from the blockchain network will be used.

\subsection{\emph{Models}}

\label{s:predictions.1}

The models chosen for our prediction-based test are Random Forest and, following \cite{nabipour2020deep,jagannath2021chain}, a special type of Recurrent Neuron Network (RNN) called long short-term memory (LSTM). With this selection we use two currently very popular predictive models that are mathematically completely different from each other. Other candidates, like Gradient Boosting or an XGBoost algorithm, are in fact very similar to a Random Forest, while Support Vector Machines (SVM), Neural Networks (NN) or Convolutional Neural Networks (CNN) belong, together with RNN, to the toolbox of Machine Learning. 

As for the configuration of the models, an optimization algorithm was used to select the hyperparameters of the LSTM, and a Grid Search algorithm for the optimization of the parameters of the Random Forest. The best settings turned to be the following.

\begin{itemize} 
\item For the LSTM: input layers=4; units=200; drop=0.1; epochs=40; look back=10; batch size=20; drop=0.1; optimizer='sgd'; units=50.
\item For the Random Forest: max features=None; max depth=8; min samples leaf=20; min samples split=30; n estimators=1100; bootstrap=False.
\end{itemize}

\subsection{\emph{Tests}}

\label{s:predictions.2}

To compare the performance of the blockchain metrics with Random Forest and LSTM, two tests were carried out, the difference being the handling of the inputs and target.

In the first test (\textit{Test 1}), the values of the blockchain metrics and target were fed in the models unaltered.

In the second test (\textit{Test 2}) and with the aim of providing the models with more generalized data, the values of the blockchain metrics and target were processed as follows: (i) the ribbons were calculated and divided by the value of the SMA-60 (the moving average of 60 days); (ii) the derivatives of the SMA-60 were calculated; (iii) the percentage of increase or decrease of the target with respect to the Bitcoin closing price was calculated.

The target for the above tests was the closing Bitcoin price 10 days after each daily record. For this target, regression models have been implemented. Therefore, the quality of the models was measured by the Root Mean Square Error (RMSE) and the Mean Absolute Scaled Error (MASE), which  compares the error of the predicted series, against a naïve prediction (the current value in the case of random walks). In addition, to check the overfitting of the models, measurements were made both in test data (20\% of the data) and in training data (80\% of the data). The corresponding Python codes are available at \cite{JCKingGithub}.

\subsection{\emph{Results}}

\label{s:predictions.3}

The results are summarized in Table \ref{table7}. In view this table, it seems that, when introducing the raw data without preprocessing, the Random Forest algorithm memorizes the values of the variables instead of interpreting them, resulting in over-fitting. Switching to mathematical textures via ribbons and mathematical derivatives seems to improve the quality of the results a bit. Furthermore, the lowest values of the MASE metric are obtained
with the algorithm \textquotedblleft LSTM with percentages\textquotedblright
, namely: 0.74 using testing data and 0.63 using training data. Ideally, the
metric should be the same in both Test and Train, but in practice this is
very difficult to achieve. However, those close values indicate that the
algorithm is interpreting and learning the patterns instead of memorizing
training data.

\begin{table}[h]
\centering
\small
\footnotesize
\begin{tabular}{|l|c|c|c|c|}
\hline
\textbf {MODEL} & \textbf {RMSE TEST} & \textbf {RMSE TRAIN} &\textbf {MASE TEST} &\textbf {MASE TRAIN} \\
\hline
Random Forest & 12229.08 & 1478.09 & 1.60 & 0.47 \\
LSTM & 5310.46 & 3555.00 & 3.32 & 9.34 \\
Random Forest with percentages & 16.30 & 9.71 & 0.82 & 0.72 \\
LSTM with percentages & 10.26 & 12.41 & 0.74 & 0.63 \\
\hline
\end{tabular}
\caption{RMSE and MASE of the two tests (with raw data and percentages) using Random Forest and LSTM.}
\label{table7}
\end{table}

It is remarkable that, other things being equal, LSTM networks obtain much better results, especially when the variables are introduced in the form of mathematical textures. In fact, it is a positive point that the difference between the RMSE of the training and test data is so small in the second test.

The metrics RMSE and MASE in Test and Train, Table\ref{table7}, cannot be compared with previous metrics in the reference \cite{jagannath2021chain} because the price of Bitcoin is much higher than that of Ethereum and the MASE is not measured in the aforementioned paper.

Another important aspect worth noting is the importance of the variables in the Random Forest model. The importance of a variable is the average of the entropies over all decision trees of a feature. The results are presented in Table \ref{table8} for raw data (Test 1), as well as in Table \ref{table9} for preprocessed data (Test 2).

The above comparison shows that, by introducing the mathematical textures of the indicators, the model is using much fewer features, but the features it uses are the ones previously seen to have a good correlation with the Bitcoin Price, like CPTRA, MWNUS, DIFF, etc.

Another lesson from our numerical experiment is that the metric AdCPTRA, that we built in Section \ref{s:applications.1.1}, is the most important for Test 2.

\begin{table}[htbp]
 
  	\centering
	\begin{tabular}{|l|c||l|c|}

      	\hline
	
	\textbf{Indicator}			& 
      	\textbf{Importance}		&
	\textbf{Indicator}			& 
      	\textbf{Importance}		 \\
	\hline
        BLCHS sma 30      & 0.1094 &TOUTV sma 60      & 0.0005 \tabularnewline  
        MKPRU sma 30      & 0.1058 &TRFUS sma 60      & 0.0004 \tabularnewline  
        MKPRU sma 30      & 0.1036 &ATRCT sma 30      & 0.0003 \tabularnewline  
        MKTCP sma 60      & 0.1003 &MWNUS diff sma 30 & 0.0003 \tabularnewline  
        BLCHS sma 60      & 0.0978 &DIFF sma 30       & 0.0002 \tabularnewline  
        CLOSE PRICE       & 0.0955 &ETRAV sma 60      & 0.0002 \tabularnewline  
        MKTCP sma 30      & 0.0951 &NTRBL sma 30      & 0.0002 \tabularnewline  
        NADDU sma 60      & 0.0432 &TRFEE sma 30 diff & 0.0002 \tabularnewline  
        ATRCT sma 60      & 0.0144 &CPTRA sma 30      & 0.0001 \tabularnewline  
        HRATE sma 60      & 0.0053 &CPTRA sma 60      & 0.0001 \tabularnewline  
        HRATE sma 30      & 0.0038 &DIFF sma 60       & 0.0001 \tabularnewline  
        RIBBON ETRAV      & 0.0023 &MIREV sma 30      & 0.0001 \tabularnewline  
        RIBBON TRFEE      & 0.0020 &MIREV sma 60      & 0.0001 \tabularnewline  
        BLCHS sma 60 diff & 0.0011 &NTREP sma 30      & 0.0001 \tabularnewline  
        BLCHS sma 60 diff & 0.0011 &RIBBON ATRCT      & 0.0001 \tabularnewline  
        NTREP sma 30      & 0.0007 &AdCPTRA           & 0.0001 \tabularnewline  

       \hline
	
	\end{tabular}
		
	\caption{Importance of variables in Random Forest, Test 1}

    \label{table8}

\end{table}

\begin{table}[htbp]
\centering
    \begin{tabular}{|l|c|} 
    \hline
    \textbf{Indicator} & \textbf{Importance} \\ \hline
    AdCPTRA & 0.3004 \\ 
    Percentage ribbon vs CPTRA sma 60 & 0.1049 \\ 
    MWNUS diff1 sma 10 & 0.0892 \\ 
    DIFF sma 60 diff1 & 0.0821 \\ 
    MKPRU sma 30 diff1 & 0.0568 \\ 
    Percentage ribbon vs ETRAV sma 60 & 0.0542 \\ 
    MKPRU sma 60 diff1 & 0.0456 \\ 
    Percentage ribbon vs NTRBL sma 60 & 0.0358 \\ 
    Percentage ribbon vs AVBLS sma 60 & 0.0309 \\ 
    RIBBON MWNUS diff1 10 20 & 0.0302 \\ 
    BLCHS sma 30 diff1 & 0.0288 \\ 
    Percentage ribbon vs BLCHS sma 60 & 0.0263 \\ 
    CPTRA sma 30 diff1 & 0.0241 \\ 
    DIFF sma 30 diff1 & 0.0138 \\ 
    NTRAT diff1 & 0.0084 \\ 
    Percentage ribbon vs TOUTV sma 60 & 0.0025 \\ 
    NTRAT diff1 sma 10 & 0.0020 \\ 
 
    \hline
    \end{tabular}
\caption{Importance of variables in Random Forest. Test 2} 
\label{table9}
\end{table}

\section{Discussion}

\label{s:discussion}

In the previous sections we have shown that the data from the blockchain network can become fundamental (MWNUS, etc) or technical (DIFF, CPTRA, etc) indicators for trading with cryptocurrencies. Indeed, the results obtained for long and short signals based on the ribbon technique demonstrate a statistical advantage. In addition, other technical indicators should be analyzed in combination with these indicators.

It is also not advisable to use a single indicator. It is more effective to use several indicators to decide to open a long or short position, or implement predictive models and operate in view of the probabilities obtained.

The mathematical methods used in this paper are mainly simple moving averages, in addition to derivatives and linear regressions in a few \textquotedblleft adjusted\textquotedblright\ indicators. Other possible methods include exponential moving averages.

Regarding the type of signals, our results with the blockchain indicators show a good performance for the long signals but not for the short signals. This indicates that blockchain indicators help to locate buy signals, but they are not very effective when it comes to locate sell signals. Therefore, it is necessary to combine this information with other information, such as that obtained directly from historical prices of the asset.

Regarding the tests with predictive models, we may conclude that LSTM networks are efficient for this type of problem, but other tests with other objectives would be necessary to confirm that the predictions are good enough to implement monitoring systems. Such tests could be buy-sell operations using the output of the predictive models or using more advanced objectives that allow to see if the price of the prediction reached the direction of the price.

\section{Conclusion}

\label{s:conclusion}

In this paper, we tested the hypothesis that blockchain indicators can help cryptocurrency traders improve the performance of their trading strategies. Nonlinear functional dependency was applied to confirm correlation with the direction of the price, while moving averages of 30 and 60 days were applied to obtain long and short signals based on the crossings of the moving averages. Furthermore, some calculations and strategies were used to measure the performance of our blockchain indicators. The datasets and Python codes used in our numerical simulations can be found in \cite{URLQuandl,JCKingGithub}.

Our results indicate that most blockchain metrics with high coefficients of
correlation with the Bitcoin price also perform well in algorithmic trading
or are variables with high entropies in predictive models. For instance,
DIFF, MKTCP and NADDU are among the first positions in functional dependence
with the Bitcoin price (Table \ref{table1}) as well
as in algorithmic trading (Tables \ref{table2}-\ref{table6}) and/or predictive models (Tables 
\ref{table8}-\ref{table9}).
But there are exceptions to this general rule. For example, ETRAV has the
lowest coefficient of correlation with the Bitcoin price and yet performs
well in algorithmic trading and is also an important variable in predictive
models.

In addition, four more advanced indicators were derived: (i) the Adjusted CPTRA ribbon (AdCPTRA), for which we applied linear regression to the historical maxima and minima of the blockchain metric CPTRA (Cost Per Transaction) to improve the results; and (ii) the Adjusted MWNUS ribbon (AdMWNUS), the Adjusted NTRAT ribbon (AdNTRAT) and the the Adjusted BLCHS ribbon (AdBLCHS), for which we used derivatives of the corresponding blockchain metric (My Wallet Number of User, Total Number of Transactions, and Blockchain API Size) to locate their most pronounced changes. Of these indicators, it is worth highlighting the AdCPTRA indicator, which is the one that obtained the best performance in short signals in algorithmic trading systems and also the most important variable in the Random Forest algorithm, exceeding the value of the second most important variable by a factor of 3. The indicators AdNTRAT and AdMWNUS had an acceptable performance and were important variables in the predictive models carried out with Random Forest, while the indicator AdBLCHS had a negative performance, although it was an important variable in both models.

Our results also show that the 18 indicators listed in Tables \ref{table3} and \ref {table4} perform good in long operations while it is risky to open short positions based on them. Both in the long and short term, better results are obtained with some indicators other than MKPRU, the indicator that uses Bitcoin prices. As for the adjusted indicators, AdCPTRA performed well both in the buy and sell simulations.

Regarding specifically the analysis of short operations, the only indicator that outperforms MKPRU is MKTCP. However, their values are very low compared to the signals in long. We conclude that short signals do not offer a statistical advantage to create a winning trading strategy.  

Moreover, some prediction tests were carried out using blockchain metrics as variables for a 7-day prediction horizon, with the LSTM networks performing best. We also observed that converting the variables to percentages helps interpret patterns and reduces overfitting. This time, the AdCPTRA indicator appears in first position in the importance of the variables with the Random Forest algorithm.

The simulation of algorithmic trading with all available blockchain ribbons also showed that blockchain metrics can outperform other metrics with higher (linear or nonlinear) correlations with the Bitcoin price (MKPRU). Such is the case of NADDU, CPTRA, TRFUS and TRFEE (Section 4.1).

We would like also to mention that our work originated from previous work of some independent analysts and traders, who publish their strategies and indicators on the TradingView portal, as in the case of Charles Edwards. We had to resort to this portal for information since it seems that nothing has been published in journals.

As a summary of this paper, it was shown that blockchain
indicators allow obtaining information with a statistical advantage in the
high volatility cryptocurrency market. Therefore, our results support that
the cryptocurrency market is adaptive in the sense of the AMH hypothesis,
that is, the data obtained from the blockchain network, although public, can
provide sufficient information to obtain profitability in the market. However, the information provided by these indicators should be combined with, e.g., trend strategies or mean reversion strategies \cite{kaufman2011alpha} when designing a trading strategy. In particular, when implementing algorithmic trading systems for cryptocurrencies based on Machine Learning or Deep Learning algorithms \cite{fister2019deep,carta2021multi}, we recommend choosing all the indicators that have performed well in any of the three analyzes carried out in Sections \ref{s:methods.4} and \ref{s:predictions} and discard only those that had a poor performance in all of them, such as the ATRCT.

\section{Acknowledgements}

We thank our Reviewers very much for their constructive criticism, which helped improve this paper.
J.M. Amig\'o and R. Dale were financially supported by Agencia Estatal de Investigaci\'on, Spain, grant PID2019-108654GB-I00/AEI/10.13039/501100011033.
J.M. Amig\'o and R. Dale were also supported by Generalitat Valenciana, Spain, grant PROMETEO/2021/063 and CIAICO/2022/252, respectively.

\bigskip

\bibliographystyle{elsarticle-num}
\bibliography{1_bibliography.bib}

\bigskip

\end{document}